\documentclass{aa}  

\usepackage{graphicx}
\usepackage{txfonts}
\usepackage{xcolor}
\usepackage{subcaption}
\usepackage{float}
\usepackage{soul}

\def\Msun{\rm M_\odot}
\def\msun{{\Msun}}

\def\Msun{\rm M_\odot}

\def\HI{\hbox{H~$\scriptstyle\rm I\ $}} 

\newcommand{\1}{\left( }
\newcommand{\2}{\right) }
\DeclareMathSymbol{*}{\mathbin}{symbols}{"01}

\begin{document} 

   \title{Astraeus IX: Impact of an evolving stellar initial mass function on early galaxies and reionisation}

   \author{Elie~R.~Cueto
          \inst{1}\thanks{\email{vdm981@alumni.ku.dk}},
          Anne~Hutter\inst{1,2}\fnmsep\thanks{\email{anne.hutter@nbi.ku.dk}},
          Pratika~Dayal\inst{3},
          Stefan~Gottl\"ober\inst{4},
          Kasper~E.~Heintz\inst{1,2},
          Charlotte~Mason\inst{1,2},
          Maxime~Trebitsch\inst{3},
          Gustavo~Yepes\inst{5,6}
          }
    \authorrunning{Cueto et al.}

   \institute{Niels Bohr Institute, University of Copenhagen, Jagtvej 128, DK-2200, Copenhagen N, Denmark
         \and
             Cosmic Dawn Center (DAWN)
         \and Kapteyn Astronomical Institute, University of Groningen, P.O. Box 800, 9700 AV Groningen, The Netherlands
         \and Leibniz-Institut f\"ur Astrophysik, An der Sternwarte 16, 14482 Potsdam, Germany
         \and Departamento de Fisica Teorica, Modulo 8, Facultad de Ciencias, Universidad Autonoma de Madrid, 28049 Madrid, Spain
         \and CIAFF, Facultad de Ciencias, Universidad Autonoma de Madrid, 28049 Madrid, Spain
             }

   \date{Received December 19, 2023; accepted - -, -}

 
    \abstract
    {Observations with the James Webb Space Telescope (JWST) have revealed an abundance of bright $z>10$ galaxy candidates, challenging the predictions of most theoretical models at high redshifts.}
    {Since massive stars dominate the observable ultraviolet (UV) emission, we explore whether a stellar initial mass function (IMF) that becomes increasingly top-heavy towards higher redshifts and lower gas-phase metallicities results in a higher abundance of bright objects in the early universe and how it influences the evolution of galaxy properties compared to a constant Salpeter IMF.}
    {We parameterised the IMF based on the findings from hydrodynamical simulations that track the formation of stars in differently metal-enriched gas clouds in the presence of the cosmic microwave background (CMB) at different redshifts. We incorporated this evolving IMF into the {\sc astraeus} (semi-numerical rAdiative tranSfer coupling of galaxy formaTion and Reionisation in N-body dArk mattEr simUlationS) framework, which couples galaxy evolution and reionisation in the first billion years. Our implementation accounts for the IMF dependence of supernova (SN) feedback, metal enrichment, and ionising and UV radiation emission. We conducted two simulations: one with a Salpeter IMF and the other with the evolving IMF. In both, we adjusted the free model parameters to reproduce key observables.}
    {Compared to a constant Salpeter IMF, we find that (i) the higher abundance of massive stars in the evolving IMF results in more light per unit stellar mass, resulting in a slower build-up of the stellar mass and lower stellar-to-halo mass ratio; (ii) due to the self-similar growth of the underlying dark matter (DM) halos, the evolving IMF's star formation main sequence scarcely deviates from that of the Salpeter IMF; (iii) the evolving IMF's stellar mass to gas-phase metallicity relation shifts to higher metallicities, while its halo mass to gas-phase metallicity relation remains unchanged; (iv) the evolving IMF's median dust-to-metal mass ratio is lower due to its stronger SN feedback; and (v) the evolving IMF requires lower values of the escape fraction of ionising photons and exhibits a flatter median relation and smaller scatter between the ionising photons emerging from galaxies and the halo mass. However, the ionising emissivities of the galaxies mainly driving reionisation ($M_h\sim10^{10}\msun$) are comparable to those of a Salpeter IMF, resulting in minimal changes to the topology of the ionised regions.}
    {These results suggest that a top-heavier IMF alone is unlikely to explain the higher abundance of bright $z>10$ sources, since the lower mass-to-light ratio driven by the greater abundance of massive stars is counteracted by stronger stellar feedback.}


   \maketitle

\section{Introduction}

The James Webb Space Telescope (JWST) plays a pivotal role in advancing our understanding of the high-redshift Universe. Its observation of galaxies at $z\gtrsim9$ in unprecedented numbers offers a unique opportunity to unravel the properties of galaxies emerging in the first few hundred million years, thereby enhancing our knowledge and constraints on galaxy evolution and reionisation. Early release observations have given us glimpses into the properties of these galaxies, painting an initial picture of compact, metal-poor, young, star-forming galaxies with stellar populations emitting harder ionising radiation \citep[e.g.][]{Bradley2023, Bunker2023, CurtisLake2023, Heintz2023a, Schaerer2022, Vanzella2023}. However, preliminary inferences based on photometric data suggest a higher abundance of UV bright galaxies than predicted by most theoretical models \citep{Labbe2023, Adams2023, Atek2023, Austin2023, BoylanKolchin2023} and that the corresponding UV luminosity functions evolves only mildly at $z>9$ \citep{Adams2023, Castellano2023, Donnan2023a, Finkelstein2023, Harikane2023a, Harikane2023c, Naidu2022}. 

These discoveries have sparked theoretical investigations into potential selection biases and modifications in physical processes that could account for the observed abundances. Some studies have proposed that the observed galaxies might be biased samples, exclusively tracing the densest regions containing the most massive galaxies \citep{McCaffrey2023} or detecting galaxies undergoing periods of intense star formation in their bursty star formation histories \citep{Mason2023, Sun2023}. While different simulations and observations agree on the presence of bursty star formation in early galaxies \citep[e.g.][and references therein]{Legrand2022, Gelli2023, Ciesla2023}, it remains unclear whether bursty star formation alone can explain the high abundance of bright $z>9$ galaxies \citep[e.g.][]{Sun2023, Pallottini2023}.
Conversely, others have explored the altered physical conditions in these very high-redshift galaxies. One possibility to reproduce the observed UV luminosity function at $z>9$ could be the ejection of dust from the star formation site through radiatively driven outflows during the initial phases of galaxy formation, such that the reduced dust attenuation compensates for the increasing shortage of bright galaxies predicted in standard theoretical models at higher redshifts \citep{Ferrara2023, Fiore2023, Mauerhofer2023, Yung2023, Ziparo2023}.
Another explanation for these UV-bright objects could be that early galaxies exhibited higher star formation efficiencies. For example, \citet{Dekel2023} estimated that if the gas in the most massive early galaxies is sufficiently dense and metal-poor, the free-fall time becomes shorter than the time required for low-metallicity stars to develop winds and SNe, resulting in feedback-free starbursts. Similarly, weaker stellar winds and fewer SNe, typical for low-metallicity stars, could weaken and delay the onset of mechanical feedback to around $\sim10$~Myr \citep{Jecmen2023, Yung2023}. An alternative explanation could involve the UV luminosity produced by massive black holes ($\gtrsim10^8\msun$) that reside in UV-bright galaxies and accrete at or slightly above the Eddington rate quasars \citep{Pacucci2022}.

Alternatively, early galaxies could have a lower mass-to-light ratio due to a higher abundance of massive stars, implying a more top-heavy IMF as suggested in \citet{Haslbauer2022, Trinca2023, Woodrum2023, Harikane2023a, Harikane2023c, Finkelstein2023, Inayoshi2022, Pacucci2022}. A more top-heavy IMF would also lead to higher ionising emissivities that would not only increase the strength of emission lines but also impact the morphology of the reionisation process of the intergalactic medium (IGM). 

Indeed, simulations of the first (metal-free) stars predominantly result in stellar masses of $\gtrsim60\msun$ \citep{Abel2002, Bromm2002, Yoshida2006, Fukushima2020}, hinting that the fraction of massive stars may rise with increasing gas temperature of star-forming clouds. 
The gas temperature not only determines whether a region within a cloud will experience gravitational collapse but also influences the presence of substructure, which, in turn, regulates the amount of material available for accretion during the collapse and subsequent formation of the star \citep[c.f. results in][]{Schneider2010, Chon2022, Sharda2022}. This dependence of the IMF on the gas temperature aligns with the results obtained from hydrodynamical simulations, showing that the IMFs of star-forming clouds become more top-heavy with a decreasing metallicity of the gas and increase in the background radiation intensity, such as the cosmic microwave background (CMB), \citep{Chon2022}. 
Indeed, the spectra of two Lyman-$\alpha$ emitters at $z\geq5.9$ \citep{Cameron2023}, along with the putative detections of Population III stellar populations at $z=6.6$ and $z=10.6$ characterised by the absence of metal lines and either strong He\,{\sc ii} or Lyman-$\alpha$, H$\gamma$, H$\beta,$ and H$\alpha$ lines \citep{Maiolino2023, Vanzella2023}, all observed with JWST, collectively argue for top-heavy IMFs. 
Furthermore, applying gas temperature-dependent IMF models to galaxy observations, where the temperature is treated as an additional parameter in the photometric template fitting, reveals that most star-forming galaxies at a fixed redshift have similar, but top-heavier IMFs than the Milky Way, with inferred gas temperatures increasing towards higher redshifts \citep{Sneppen2022, Steinhardt2022a, Steinhardt2022b}.
In addition, observations of local star-forming regions suggest that the IMF varies across correlated star formation events. For instance, studies of globular clusters, ultra-compact dwarf galaxies and young massive clusters indicate that the IMF becomes top-heavy in low metallicity and dense gas environments \citep[e.g.][]{Dabringhausen2009, Dabringhausen2012, Marks2012, Zonoozi2016, Haghi2017, Kalari2018, Schneider2018, Dib2023}. In contrast, it becomes more bottom-heavy in metal-rich environments \citep[e.g.][]{Marks2012, Chabrier2014}, such as in the centres of nearby elliptical galaxies \citep[e.g.][]{vanDokkumConroy2010, Conroy2017}. Averaging the IMF over all star-forming regions in a galaxy, this galaxy-wide IMF (gwIMF) \citep{Kroupa2003, Weidner2013, Yan2017, Jerabkova2018} is often found to be top-heavy in galaxies with high star formation rates and top-light in galaxies with low star formation rates \citep[e.g.][]{Meurer2009, Watts2018, Zhang2018, Gunawardhana2011, Fontanot2017, Fontanot2018a}.

The above-quoted studies highlight that the IMF in early galaxies is very likely to differ from the canonical Milky Way and present-day IMF that is used in nearly all state-of-the-art large-scale reionisation radiation hydrodynamics and semi-numerical simulations following the evolution of galaxies explicitly \citep{Ocvirk2020, Lewis2022, Kannan2022, Gnedin2014, Mutch2016, Hutter2021a}. To address this discrepancy, some simulations and semi-analytic models of galaxy evolution have introduced a distinction between the top-heavy IMF associated with the first stars (Population III) and the present-day canonical IMF for metal-enriched Population II stars \citep[e.g.][]{Maio2010, Norman2018, Visbal2020}. However, these do not consider the IMF of Population II stars to evolve or depend on the galaxies' properties. Only at lower redshifts, a few hydrodynamics simulations and semi-analytic galaxy evolution models have incorporated and explored the integrated galaxy-wide IMF model \citep{Weidner2005, Weidner2013} (assuming that the most massive star in a cluster is linked to its cluster mass) to explore the impact of an empirically informed varying IMF on the evolution of galaxy properties \citep{Ploeckinger2014, Fontanot2017}. More recently and at higher redshifts, only \citet{Trinca2023} have explored the effect of an IMF varying with stellar metallicity and redshift on the galaxy UV luminosity functions, finding the resulting UV LFs to better fit the observations at $z>9$. However, these authors considered this evolving IMF only when deriving the galaxies' UV luminosities in their semi-analytic galaxy evolution model, not when evaluating stellar feedback and metal yields for determining galaxy properties or the ionising photon production for reionisation.

Omitting the IMF's dependency on stellar feedback can lead to severely erroneous conclusions. Firstly, a more top-heavy IMF will not only reduce the mass-to-light ratio, but also enhance the fraction of stars exploding as SNe, thereby reducing subsequent star formation more immediately while increasing the metal enrichment of the interstellar gas per exploding SN. Given the theoretical and observational hints towards an evolving IMF, we introduce the first model that follows the mutual evolution of galaxies and reionisation, while assuming an IMF that evolves in each galaxy according to the metallicity of its star-forming gas and redshift. For this purpose, we employed a parameterisation of the evolving IMF that follows the results from the spherical turbulent gas cloud simulations in the presence of the CMB at various gas metallicities and redshifts \citep{Chon2022}. We incorporated this IMF parameterisation into our {\sc astraeus} framework, a semi-numerical model that tracks the interdependent evolution of galaxies and reionisation \citep{Hutter2021a, Ucci2023, Hutter2023a} by incorporating IMF-dependence into all relevant physical processes.
With this updated model, we investigate the following question during the Epoch of Reionisation (EoR), which spans between $z\simeq5-15$ and  during which most of the intergalactic medium (IGM) was ionised at $z\lesssim7.5$ \citep{planck2018}: How does an evolving IMF affect the properties of early galaxies and reionisation, and can it explain the evolution of the observed UV LFs during the EoR? 
This paper is organised as follows. In Sect. \ref{sec_model}, we briefly describe the {\sc astraeus} model and outline how we parameterise and integrate an evolving IMF into {\sc astraeus}. Sect. \ref{sec_baselining_model_against_observations} describes how we tune the free {\sc astraeus} model parameters to reproduce the observed UV luminosity functions and reionisation history constraints for both the constant Salpeter IMF and our new evolving IMF scenarios. We then discuss how an evolving IMF changes the mass-to-UV light ratio, stellar-to-halo mass ratio, star formation sequence, stellar mass to metallicity and stellar mass to dust mass relations, as well as the UV luminosity to metallicity and UV luminosity to dust mass relations, compared to a constant Salpeter IMF (Sect. \ref{sec_results}). 
We present our conclusions in Sect. \ref{sec_conclusions}. In this paper, we assume the AB magnitude system \citep{Oke1983} and a $\Lambda$CDM universe with the following \citet{planck2016} cosmological parameters:  $\Omega_\Lambda=0.692885$, $\Omega_m=0.307115$, $\Omega_b=0.048206$, $H_0=100h=67.77$km~s$^{-1}$Mpc$^{-1}$, $n_s=0.96$, and $\sigma_8=0.8228$.

\section{The model }
\label{sec_model}

The {\sc astraeus} framework couples an enhanced version of the semi-analytic galaxy evolution model {\sc delphi} \citep{dayal2014, Dayal2022} with the semi-numerical reionisation scheme {\sc cifog} \citep{Hutter2018a}. The resulting model runs on the outputs of a dark matter (DM) only N-body simulation. In this section, we briefly revisit the physical processes tracked in {\sc astraeus} and described in \citet{Hutter2021a, Ucci2023} and \citet{Hutter2023a} in detail. Here, we also describe our implementation of an evolving IMF. 

\subsection{The N-body simulation}

We used the {\sc vsmdpl} (very small multidark planck) DM-only N-body simulation, which is part of the Multidark simulation project\footnote{\url{http://www.cosmosim.org/}} and has been run with the {\sc gadget-2 tree+pm} code \citep{springel2005}. The simulation encompasses a cubic volume with a side length of $160h^{-1}$~comoving Mpc (cMpc) and tracks the trajectories of $3840^3$ DM particles. Each DM particle carries a mass of $6\times10^6h^{-1}\Msun$. For 150 snapshots spanning from $z=25$ to $z=0$, halos and subhalos down to 20 particles or a minimum mass of $1.24\times10^8h^{-1}\msun$ have been identified with the phase space {\sc rockstar} halo finder \citep{behroozi2013_rs}. Since {\sc astraeus} includes time-synchronised processes like reionisation, we have used the pipeline internal {\sc cutnresort} scheme to re-sort the vertical merger trees generated by {\sc consistent trees} \citep{behroozi2013_trees} to local horizontal merger trees for all galaxies at $z=4.5$ \citep[for details see Appendix A in][]{Hutter2021a}. For the first 74 snapshots spanning from $z=25$ to $z=4.5$, we generate the DM density fields required as input for the {\sc astraeus} pipeline by mapping the DM particles onto $2048^3$ grids and subsequently resampling to these $512^3$ grids.

\subsection{Galaxy evolution}

{\sc astraeus}\footnote{\citet{astraeus}, \url{https://github.com/annehutter/astraeus}} follows the key physical processes of early galaxy formation and reionisation. By post-processing the DM merger trees and density fields from the {\sc vsmdpl} simulation, it tracks for each galaxy and at each time step (i.e. snapshot of the N-body simulation) the amount of gas accreted, the gas and stellar mass merged, the formation of stars and their feedback through SNe and metal enrichment, as well as the large-scale reionisation process and its feedback on the gas content in galaxies. The modelling of these processes is described below.

\subsubsection{Gas and stars}

Each galaxy that starts forming stars in a halo with mass $M_h$, is assumed to have a gas mass of $M_g^i(z)=f_g (\Omega_b/\Omega_m) M_h(z)$. Here, $f_g$ describes the gas fraction that is not evaporated by reionisation, assuming values of $f_g<1$ and $f_g=1$ as the galaxy forms in an ionised and neutral region, respectively.\footnote{The exact value of $f_g$ depends on the assumed radiative feedback model as outlined in \citet{Hutter2021a}. It decreases, the earlier the environment of a galaxy has been ionised, the higher the incident photoionisation rate or heating of the ionised gas is.} In subsequent time steps, the gas mass of the galaxy includes the gas inherited from its progenitor, $M_g^\mathrm{mer}(z)$ and gained by smooth accretion, $M_g^\mathrm{acc}(z)$, but never exceeds the limit given by reionisation feedback \citep{Gnedin2000, Sobacchi2013}:
\begin{eqnarray}
M_g^i(z) &=& \min\left( M_g^\mathrm{mer}(z) + M_g^\mathrm{acc}(z), f_g (\Omega_b/\Omega_m) M_h \right),
\end{eqnarray}
with 
\begin{eqnarray}
M_g^\mathrm{acc}(z) &=& \frac{\Omega_b}{\Omega_m} \left[ M_h(z) - \sum_{p=1}^\mathrm{N_p} M_{h,p}(z + \Delta z) \right] ,\\
M_g^\mathrm{mer}(z) &=& \sum_{p=1}^\mathrm{N_p} M_{g,p} (z + \Delta z).
\end{eqnarray}
Here, $N_p$ is the number of progenitors of a galaxy, while $M_{h,p}$ and $M_{g,p}$  are each of its progenitor's halo and gas masses at $z+\Delta z$.
At each time step, we assume that a fraction of the merged and accreted gas mass $M_g^i$ forms stars over the time step's length, $\Delta t$, amounting to a mass of newly formed stars of $M_\star^\mathrm{new}(z)=f_\star^\mathrm{eff} M_g^i(z)$. $f_\star^\mathrm{eff}$ depicts the fraction of gas that forms stars in $\Delta t$, namely, the star formation efficiency. We note that its value depends on the gravitational potential of the galaxy: In massive galaxies the fraction of gas converted into stars reaches the maximum value, given by $f_\star \frac{\tau_\mathrm{dyn}(z=9)}{\tau_\mathrm{dyn}(z)} \frac{20~\mathrm{Myrs}}{\Delta t}$, where $\tau_\mathrm{dyn}(z)=\sqrt{r_\mathrm{vir}^3/(G~M_{h})}\propto (1+z)^{-3/2}$ is the dynamic time of the system at redshift $z$.\footnote{We assume $f_\mathrm{\star}$ to be proportional to $\tau_\mathrm{dyn}^{-1}(z)$ to account for the enhanced gas density in galaxies towards higher redshifts.} However, in lower mass galaxies, SN and radiative feedback from reionisation limit this fraction further. Our model assumes that a galaxy can form only as many stars as are necessary to eject all gas from the galaxy due to SN explosions, with the corresponding fraction given as:
\begin{eqnarray}
f_\star^\mathrm{ej}(z) &=& \frac{v_c^2}{v_c^2 + f_w^\mathrm{eff}(z)\ E_{51} \nu_z} \left[  1 - \frac{f_w^\mathrm{eff}(z)\ E_{51} \sum_j \nu_j M_{\star,j}^\mathrm{new}(z_j)}{M_\mathrm{g}^i(z)~ v_c^2} \right].
\label{eq_fstarEj}
\end{eqnarray}
Here, $v_c$ is the rotational velocity of the halo, $E_{51}$ the energy released by a type II supernova (SNII), $\nu_z$ the IMF-dependent fraction of stellar mass forming and exploding in the current time step, and
\begin{eqnarray}
f_w^\mathrm{eff}(z) &=& \frac{f_w}{1 + \left( \frac{20~\mathrm{Myr}}{\Delta t} - 1\right) \frac{M_g^\mathrm{mer}(z)}{M_g^i(z)}}
\end{eqnarray}
is the fraction of SN energy injected into the winds driving gas outflows. The right term in brackets arises from our delayed SN feedback scheme, where at each time step the effective star formation efficiency accounts also for the SN energy released from stars formed in previous time steps, following the mass-dependent stellar lifetimes in \citet{padovani1993}. Thus, $M_{\star,j}^\mathrm{new}(z_j)$ is the stellar mass formed during previous time step $j$, and $\nu_j$ the fraction of stellar mass formed in previous time step $j$ that explodes in the current time step given the IMF of the respective progenitor. Thus, the fraction of gas converted into stars, effectively the star formation efficiency, is given by $f_\star^\mathrm{eff}=\min(f_\star \frac{\tau_\mathrm{dyn}(z=9)}{\tau_\mathrm{dyn}(z)} \frac{20~\mathrm{Myrs}}{\Delta t}, f_\star^\mathrm{ej})$ at each time step. Both, $f_\star$ and $f_w^\mathrm{eff}$ are free model parameters.
The total stellar mass at redshift, $z, $ is then:\ 
\begin{eqnarray}
    M_\star(z) &=& \sum_{p=1}^\mathrm{N_p} M_{\star,p} (z + \Delta z) ~ + ~ M_\star^\mathrm{new}(z) ~ - ~ M_\star^\mathrm{destr}(z),
\end{eqnarray}
where $M_\star^\mathrm{destr}(z)$ is the stellar mass that is returned into gas and metals in SNe and AGB stars.

\subsubsection{Metals and dust}

{\sc astraeus} tracks the metal enrichment by stellar winds, SN Type II (SNII) and SN Type Ia (SNIa) explosions and AGB stars \citep[for a detailed description see][]{Ucci2023}. At each time step, we assume the smoothly accreted gas to have the average metallicity of the intergalactic medium (IGM), $Z_\mathrm{IGM}$. The quantity of newly formed metals depends on the number of massive stars exploding as SNe during the current time step. We use the stellar lifetimes from \citet{padovani1993}, metal enrichment rates from stellar winds, SNII and SNIa as described in \citet{Yates2013} with the SNIa progenitor fraction and delay-time distribution as defined in \citet{Arrigoni2010} and \citet{Maoz2012}, and the most recent metal yields from \citet{Kobayashi2020b}. Furthermore, we assume the galaxy's gas and metals to be perfectly mixed. Thus, the amount of ejected metals is directly proportional to the ejected gas mass and the metallicity of the gas in the galaxy and contributes to $Z_\mathrm{IGM}$.

{\sc astraeus} also follows the formation, growth, destruction, astration, and ejection of dust in each galaxy, where our model assumes the dust mass reservoir to be part of the metal mass reservoir. Specifically, our model assumes that dust is produced by SNII and AGB stars through the condensation of metals in stellar ejecta, and its grains grow through the accretion of heavy elements in dense molecular clouds in the ISM. The main mechanisms reducing the dust mass in a galaxy are the destruction by SN blastwaves, the formation of new stars (astration) and the ejection of metal-enriched gas. All these processes account for the IMF and the varying lifetimes of stars with different masses. We refer the reader to \citet{Hutter2023a} and \citet{Dayal2022} for the exact formalism that derives the dust mass of each galaxy. 

For each galaxy, we follow \citet{Hutter2023a} to derive the attenuation of UV continuum photons ($\lambda\simeq1500$\AA) by dust from its dust mass, virial radius and spin. The dust mass, $M_d$, its distribution radius, $r_d$, and the radius, $a=0.03~\mu$m, and material density, $s=2.25$~g~cm$^{-3}$, of the graphite/carbonaceous dust grains provide an estimate for the optical depth to the UV continuum photons,
\begin{eqnarray}
    \tau_\mathrm{UV,c} &=& \frac{3\sigma_d}{4as},
\end{eqnarray}
where $\sigma_d=M_d/(\pi r_d^2)$ is the dust surface mass density. We assume the gas and dust in each galaxy to be perfectly mixed ($r_d=r_g$) where the radius of the gas is given by $r_g=4.5\lambda r_\mathrm{vir}\left[ (1+z)/6\right]^{1.8}$. Here, $\lambda$ and $r_\mathrm{vir}$ are the spin parameter and virial radius of the simulated halo, respectively, while the third factor reflects the redshift evolution of the compactness of galaxies and is motivated by the non-evolution of [C{\small II}] sizes from $z\simeq7$ to $z\simeq4.5$ \citep{Fudamoto2022}. We derive the escape fraction of the UV continuum photons by assuming a slab-like geometry,
\begin{eqnarray}
    f_\mathrm{esc}^\mathrm{c} &=& \frac{1-\exp{(-\tau_\mathrm{UV,c})}}{\tau_\mathrm{UV,c}}.
\end{eqnarray}  
The observed UV luminosity is then given by the dust attenuated intrinsic UV luminosity
\begin{eqnarray}
    L_\mathrm{c} ^\mathrm{obs} &=& f_\mathrm{esc}^\mathrm{c} L_\mathrm{c} ,
\end{eqnarray}
with the intrinsic UV luminosity $L_\mathrm{c} $ being computed as described in \citet{Hutter2021a}.

\subsection{Incorporating an evolving IMF}
\label{subsec_evolving_IMF}

\begin{figure*}
    \centering
    \includegraphics[width = \textwidth]{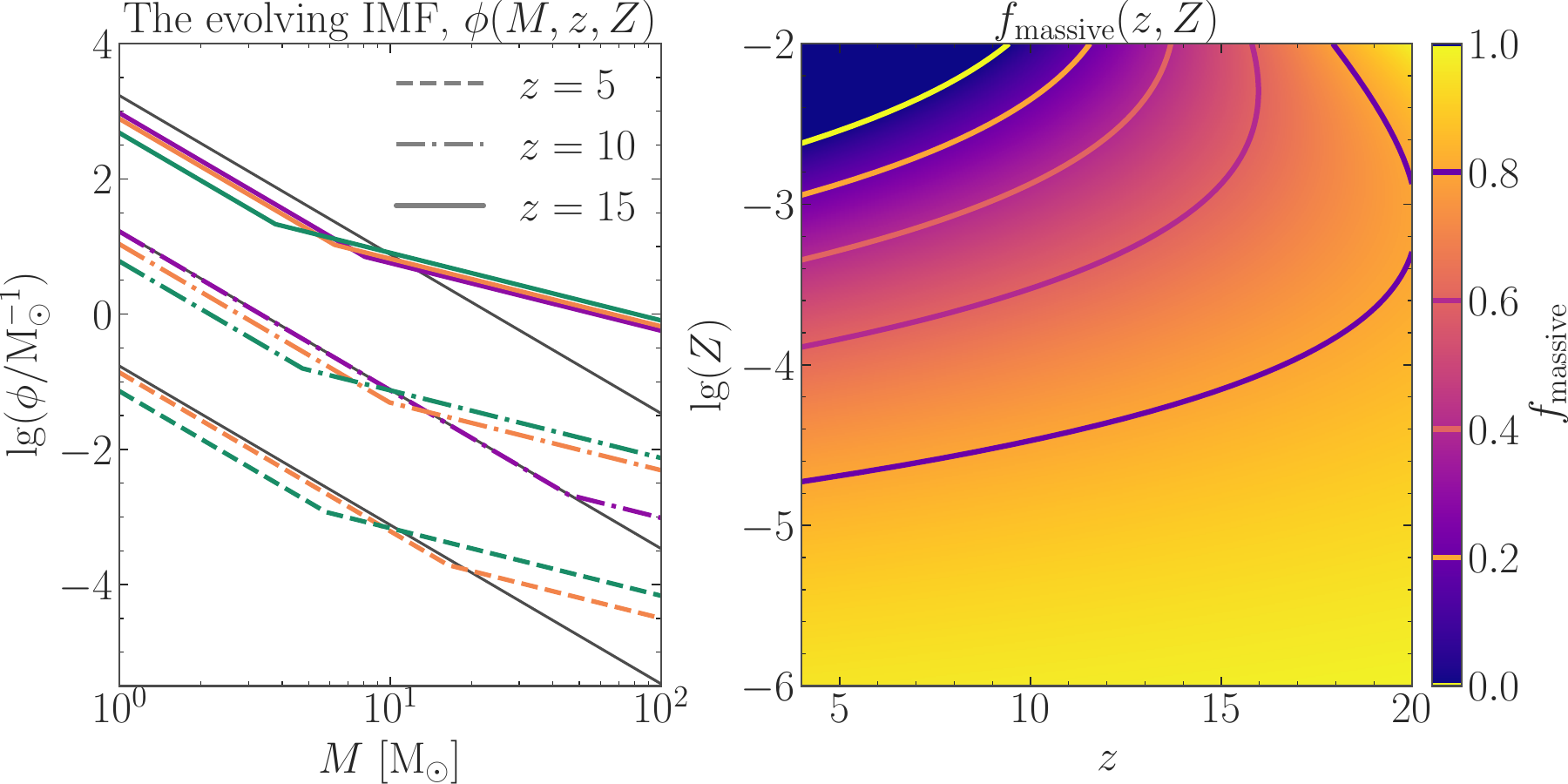}
    \caption{The evolving IMF compared with the Salpeter IMF, and the evolution of the massive fraction with redshift and metallicity. \textit{Left:}  Evolving IMF at different absolute metallicities $Z=10^{-2}$ (purple lines), $10^{-3}$ (orange lines), $10^{-4}$ (green lines) and redshifts $z=5$ (dashed lines), $10$ (dot-dashed lines), and $15$ (solid lines). For clarity, the evolving IMFs at $z=10$ and $15$ have been displaced by $\lg(\phi)=2$ and $4$, respectively. Thin black lines show the Salpeter IMF at the depicted redshifts. \textit{Right:}  Massive fraction $f_{\rm massive}$ depicted as a function of redshift $z$ and absolute metallicity $Z$. The top-left corner shows the evolving IMF developing into the Salpeter IMF, where $f_{\rm massive}=0$.}
    \label{fig:IMF}
\end{figure*}

In this work, we extended {\sc astraeus} to include an IMF that evolves with redshift and depends on the metallicity of the star-forming gas as found in the hydrodynamics simulations presented in \citet{Chon2022}. These simulations follow the collapse of a spherical turbulent gas cloud in the presence of the CMB at different redshifts and gas metallicities. They suggest that the IMF can be composed of a component with a Salpeter-like slope at the low stellar mass end and a log-flat component at the high stellar mass end, with the fraction of stellar mass produced in the log-flat component increasing with rising redshift and decreasing gas metallicity -- originating from the CMB heating the gas to higher temperatures at higher redshifts and gas cooling becoming less efficient for metal-poorer gas. By fitting the results from these simulations, \citet{Chon2022} derives the respective scaling of the fraction of stellar mass in the log-flat component as
\begin{eqnarray}
    f_\mathrm{massive} &=& 1.07 * (1-2^x) + 0.04 * 2.67^x * z,\\
    x &=& 1 + \lg\1 \frac{Z}{Z_\odot}\2,
    \label{eq_fmassive}
\end{eqnarray}
with $Z_\odot = 0.0134$ \citep{Asplund2009}, and $z$ and $Z$ being the redshift and gas metallicity of the star-forming cloud, respectively.
To efficiently implement such an evolving IMF into the {\sc astraeus} simulation framework, we define the IMF, $\phi(M)=\frac{\mathrm{d}N}{\mathrm{d}M}$, as
\begin{eqnarray}
    \frac{\mathrm{d}N}{\mathrm{d}M} = 
    \begin{cases}
        \left(1 - f_\mathrm{massive}\right)\ \frac{M_\mathrm{i}^{-0.35} - M_\mathrm{f}^{-0.35}}{M_\mathrm{i}^{-0.35}-M_\mathrm{c}^{-0.35}}\ M^{-2.35} & ,\ M_\mathrm{i} \leq M < M_\mathrm{c}\\
        f_\mathrm{massive}\ \frac{M_\mathrm{f} - M_\mathrm{i}}{M_\mathrm{f} - M_\mathrm{c}}\ M^{-1} & ,\ M_\mathrm{c} \leq M \leq M_\mathrm{f},
    \end{cases}
    \label{eq_evolvingIMF}
\end{eqnarray}
where $M_\mathrm{i}$ and $M_\mathrm{f}$ are $0.1~\msun$ and $100~\msun$ respectively, $M_\mathrm{c}$ the cutoff mass that describes the mass where the Salpeter component transitions to the log-flat component, and $f_\mathrm{massive}$ the stellar mass fraction above $M_\mathrm{c}$.\footnote{We choose this definition, which slightly differs from the one described in \citet{Chon2022}, to ensure that the IMF remains continuous.} We show how the IMF and $f_\mathrm{massive}$ depend on redshift and gas metallicity in Fig.~\ref{fig:IMF}. In practice, we derive $f_\mathrm{massive}$ from Eq. \ref{eq_fmassive}, while we infer $M_\mathrm{c}$ by solving $\int_{M_\mathrm{i}}^{M_\mathrm{f}} \phi(M)~M~\mathrm{d}M = 1\msun$ numerically. To speed up computations, we generate a corresponding look-up table for $M_\mathrm{c}$ that covers $f_\mathrm{massive}$ values ranging from $0$ to $1$ in steps of $0.001$.
We note that we choose an upper mass limit of $M_\mathrm{f}=100\msun$ to be consistent with the parameterised upper mass limit of the IMF in \citet{Chon2022} as well as previous {\sc astraeus} simulation runs.

Several physical processes implemented in {\sc astraeus}, such as SN feedback, metal enrichment and the ionising emissivity, depend on the assumed IMF. A key challenge in incorporating an evolving IMF is that these processes depend on the evolution of the IMF throughout the star formation history (SFH) of a galaxy. While our implementation of these processes can easily account for the redshift dependence of the IMF, tracking the gas metallicities of all progenitors of each galaxy would require advanced book-keeping structures in the simulation code. Therefore, for a galaxy at redshift $z$, we assume that its progenitors at redshift $z_j$ formed stars with a metallicity mass-averaged over all $N_p$ progenitors at $z_j$,
\begin{eqnarray}
    Z(z_j) &=& \frac{ \sum_{p=1}^{N_p} M_\mathrm{metal}^i(z_j) }{ \sum_{p=1}^{N_p} M_\mathrm{g}^i(z_j) }.
    \label{eq_metallicityhistory}
\end{eqnarray}
Here, $M_\mathrm{metal}^i$ and $M_\mathrm{g}^i$ are the initial metal and gas masses of the progenitor at $z_j$, respectively. While this assumption becomes less accurate as the number of progenitors increases, it allows us to assign a single metallicity history to each simulated galaxy, making the implementation of the evolving IMF straightforward. We plan to develop a more accurate representation in future work, such as storing the metallicities of the progenitors in the last $30$~Myrs.

Thus, in our delayed SN feedback scheme, the fraction of stellar mass forming and exploding in the current time step, $\nu_z$, in Eq. \ref{eq_fstarEj} is derived from the IMF that is determined by the galaxy's current redshift $z$ and gas metallicity. However, the fraction of stellar mass forming in previous time steps and exploding in the current time step, $\nu_j$, is determined by the redshift $z_j$ of the previous time step and the metallicity mass-averaged over all progenitors at $z_j$.

We also account for the evolving IMF when deriving the metal yields from SNII and SNIa explosions as well as AGB stars by accounting for the evolving IMF, $\phi(M)$, in Eq.~8 in \citet{Ucci2023}. For SNIa, we adjusted the fraction of stellar systems in the mass range $3-16~\msun$, $f_\mathrm{3-16}$, representing $2.8\%$ of SNIa progenitors, and the number of stars produced per $1~\msun$, $k$, to the evolving IMF. We adjusted the second term in the summation of Eq.~8 in \citet{Ucci2023} to $A~k~\int_{\tau(0.85\msun)}^{\tau(8\msun)}~M^\mathrm{SNIa}~f_\mathrm{3-16}(t-\tau)~\mathrm{SFR}(t-\tau)~\mathrm{DTD}(t)~\mathrm{d}\tau$ with the analytic power-law SNIa delay-time distribution $\mathrm{DTD}$ as defined in \citet{Maoz2012} and \citet{Yates2013}.

To derive the ionising emissivity and UV luminosity of each galaxy, we have used the stellar population synthesis code {\sc starburst99} \citep{Leitherer1999} to generate the spectra for a starburst assuming a Salpeter IMF between $0.1\msun$ and $M_c$ and a log-flat IMF between $M_c$ and $100~\msun$, for $M_c$ values ranging from $0-100\msun$ and metallicity values ranging from $Z=0.001$ to $0.008$. From the {\sc starburst99} simulation outputs we fit the time evolutions of the ionising emissivity and UV luminosity as a function of redshift and metallicity, with the latter two defining the IMF. For the ionising emissivity, we have:%

{\small
\begin{eqnarray}
    \lg\frac{\dot{Q}(t, z, Z)}{\mathrm{s}^{-1}\msun^{-1}} &=& 
        45.93 - 1.17~ (\lg Z+2)
        + 0.064~ z~ (\lg Z + 3.46) \nonumber \\
        &-& \mathrm{max}[0,~\lg t - 6.35] \label{eq:Q} \\ 
        &\times& \left( \left[3.91 + z((\lg Z + 2.97) * 0.051) + 0.0293\right] \right).   \nonumber
\end{eqnarray}
}%
Then, for the UV luminosity, we have:\ 
{\small
\begin{eqnarray}\label{eq:Lnu}
    \lg \frac{L_\nu(t, z, Z)}{\mathrm{erg}\mathrm{s}^{-1}\mathrm{Hz}^{-1}\msun^{-1}} &=&  
        31.94 - 0.92~(\lg Z+1.68) \\
        &+& 0.06~z~(\lg Z + 3.50) \nonumber\\
        &+& 0.69\ \mathrm{max}[0,~\mathrm{min}(6.45,\lg t)-6.0] \nonumber\\
        &-& (2.0 - 0.032~z)~ \mathrm{max}[0,~\lg t-7.32] \nonumber\\
        &-& (1.56 + 0.06~z)~ \nonumber \\
        &&\times\ \mathrm{max}[0,\mathrm{min}(7.32,~\lg t)-6.45]. \nonumber
\end{eqnarray}
}%
We note that since the ionising emissivity of a galaxy is computed on the fly, we used the galaxy's mass-averaged metallicities as defined in Eq. \ref{eq_metallicityhistory}. As we derive a galaxy's UV luminosity in post-processing, we account for the exact gas metallicities of its progenitors. We also assume that the newly formed stellar mass, $M_\star^\mathrm{new}$, is formed at a constant rate throughout the time step. This requires convolving Eqs. \ref{eq:Q} and \ref{eq:Lnu} with the respective top-heavy star formation history; we provide the resulting expressions in Appendix \ref{app_luminosities_contSF}. 

\begin{table}
  \centering
  \caption{{\sc astraeus} model parameters values for the Salpeter and evolving IMF models.}
  \label{tab:example_table}
  \begin{tabular*}{\columnwidth}{ccc}
    \hline\hline
    Parameter & Salpeter IMF & Evolving IMF \\
    \hline
    $f_\star$ & $0.025$ & $0.01$ \\
    $f_w$      & $0.2$   & $0.3$ \\
    Radiative feedback & Photoionisation & Photoionisation \\
    IMF & \citet{Salpeter} & \citet{Chon2022} \\
    SED & \textsc{Starburst99} & \textsc{Starburst99} \\
    $f_\mathrm{esc}^{0}$ & 0.31 & 0.038 \\
    $\sigma_e$ & 0.0543 & 0.0580 \\
    \hline
  \end{tabular*}
  \label{table_model_params_fsdynamic}
\end{table}

\subsection{Reionisation}
\label{subsec_reionisation}

{\sc astraeus} follows the reionisation of the IGM on the fly. At each time step, it derives the number of ionising photons generated in each galaxy by convolving the star formation and metallicity histories with the evolving ionising emissivities of the respective stellar populations (see Eq. \ref{eq:Q} for the ionising emissivities' redshift and metallicity dependencies). The number of ionising photons that escape from each galaxy to contribute to the ionisation of the IGM is given by:
\begin{eqnarray}
    \dot{N}_\mathrm{ion} &=& f_\mathrm{esc} \ \dot{Q}.
\end{eqnarray}
Here, $f_\mathrm{esc}$ represents the fraction of ionising photons produced in the galaxy that escape into the IGM. The {\sc mhdec} model in \citet{Hutter2023a} and the results in \citet{Ocvirk2021} suggest that an $f_\mathrm{esc}$ value effectively decreasing with halo mass provides a better fit to the global \HI fraction at $z<6$. For this reason, we adopt the physically motivated $f_\mathrm{esc}$ model from \citet{Hutter2021a} where $f_\mathrm{esc}$ scales with the ejected gas fraction, i.e. $f_\mathrm{esc} = f_\mathrm{esc}^0 \min(1, f_\star^\mathrm{eff} / f_\star^\mathrm{ej})$ with $f_\mathrm{esc}^0$ being a free parameter. 
Using the resulting ionising emissivities $\dot{N}_\mathrm{ion}$ of each galaxy at the time step and the distribution of the intergalactic gas density (assumed to follow the DM perfectly), {\sc astraeus} computes the spatial distribution of the ionised regions in the simulation box. To determine whether a region within the simulation box is ionised, it compares the region's cumulative number of ionising photons with its number of absorption events at different smoothing scales \citep[see][for the details of the algorithm]{Hutter2018a}, accounting in this way for the ionising radiation from neighbouring sources. Within ionised regions, {\sc astraeus} also computes the photoionisation rate and the residual neutral hydrogen (\HI) fraction, outputting photoionisation and ionisation fields. From these outputs, it determines (i) when the environment of a galaxy was reionised and (ii) the corresponding photoionisation rate. Both are then used to account for the radiative feedback from reionisation by computing the gas mass the galaxy can retain ($f_g M_g^i$). For the latter, we adopt the Photoionisation model described in \citet{Hutter2021a}, representing a weak to intermediate, time-delayed radiative feedback. In this model, a galaxy can only hold on to its gas if it exceeds a characteristic mass $M_\mathrm{char}$. In turn, $M_\mathrm{char}$ increases as the photoionisation rate $\Gamma_\mathrm{HI}$ incident at the galaxy's location at $z_\mathrm{reion}$ and/or the difference between $z_\mathrm{reion}$ and the galaxy's current redshift $z$ rises \citep[see also][]{Sobacchi2013}, where $z_\mathrm{reion}$ marks the redshift at which the environment of the galaxy became reionised.

\begin{figure*}
    \centering
    \includegraphics[width = \textwidth]{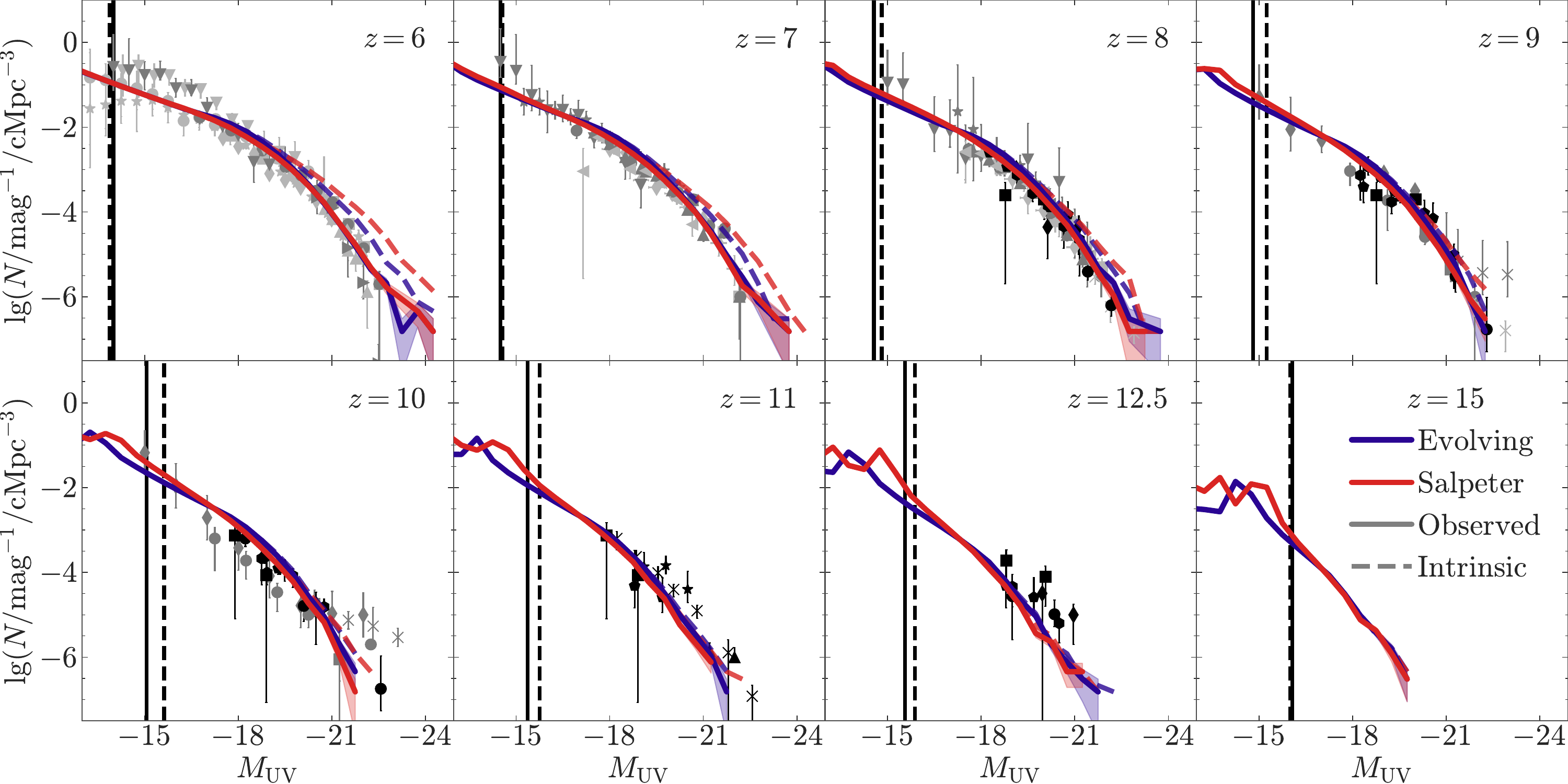}
    \caption{UV luminosity functions (LFs) at $z=6-11$, $12.5$ and $15$ for the evolving IMF (blue lines) and the Salpeter IMF (red lines). Dashed lines show the intrinsic UV LFs, while solid lines depict the dust attenuated UV LFs. Vertical black solid and dashed lines denote the UV luminosity thresholds in the evolving IMF and Salpeter IMF simulations, respectively, below which the star formation rates have not converged for all galaxies due to the mass resolution limit of the {\sc vsmdpl} simulation.
    Black points show the observational constraints from JWST data: 
    \citealt{Adams2023} (hexagons),
    \citealt{bouwens_uv_2023} (diamonds), 
    \citealt{bouwens_evolution_2023} (squares), 
    \citealt{casey_cosmos-web_2023} (triangles), 
    \citealt{mcleod_galaxy_2023} (x's), 
    \citealt{donnan_evolution_2023} (circles), 
    \citealt{Finkelstein2023} (stars), 
    \citealt{Willot2023} (pentagons). 
    Light and dark grey points depict the observational constraints before JWST: 
    \citealt{atek_new_2015} (light stars), 
    \citealt{atek_extreme_2018} (dark stars),
    \citealt{bouwens_uv_2015} (light squares), 
    \citealt{bouwens_bright_2016} (dark squares), 
    \citealt{bouwens_z_2017} (light circles),
    \citealt{bouwens_new_2021} (dark circles),
    \citealt{bowler_lack_2020} (light x's), 
    \citealt{calvi_bright_2016} (dark x's), 
    \citealt{finkelstein_evolution_2015} (light diamonds),
    \citealt{ishigaki_full-data_2018} (dark diamonds),
    \citealt{livermore_sneak_2016} (light 'downward' triangles),
    \citealt{livermore_uv_2017} (dark 'downward' triangles), 
    \citealt{mclure_luminosity_2009} (light 'upward' triangles), 
    \citealt{mclure_new_2013} (dark 'upward' triangles), 
    \citealt{schenker_uv_2013} (light 'left' triangles), 
    \citealt{schmidt_luminosity_2014} (dark 'left' triangles),
    \citealt{van_der_burg_uv_2010} (light 'right' triangles),
    \citealt{willott_exponential_2013} (dark 'right' triangles).}
    \label{fig:UVLF}
\end{figure*}

\begin{figure}
    \centering
    \includegraphics[width = 0.5\textwidth]{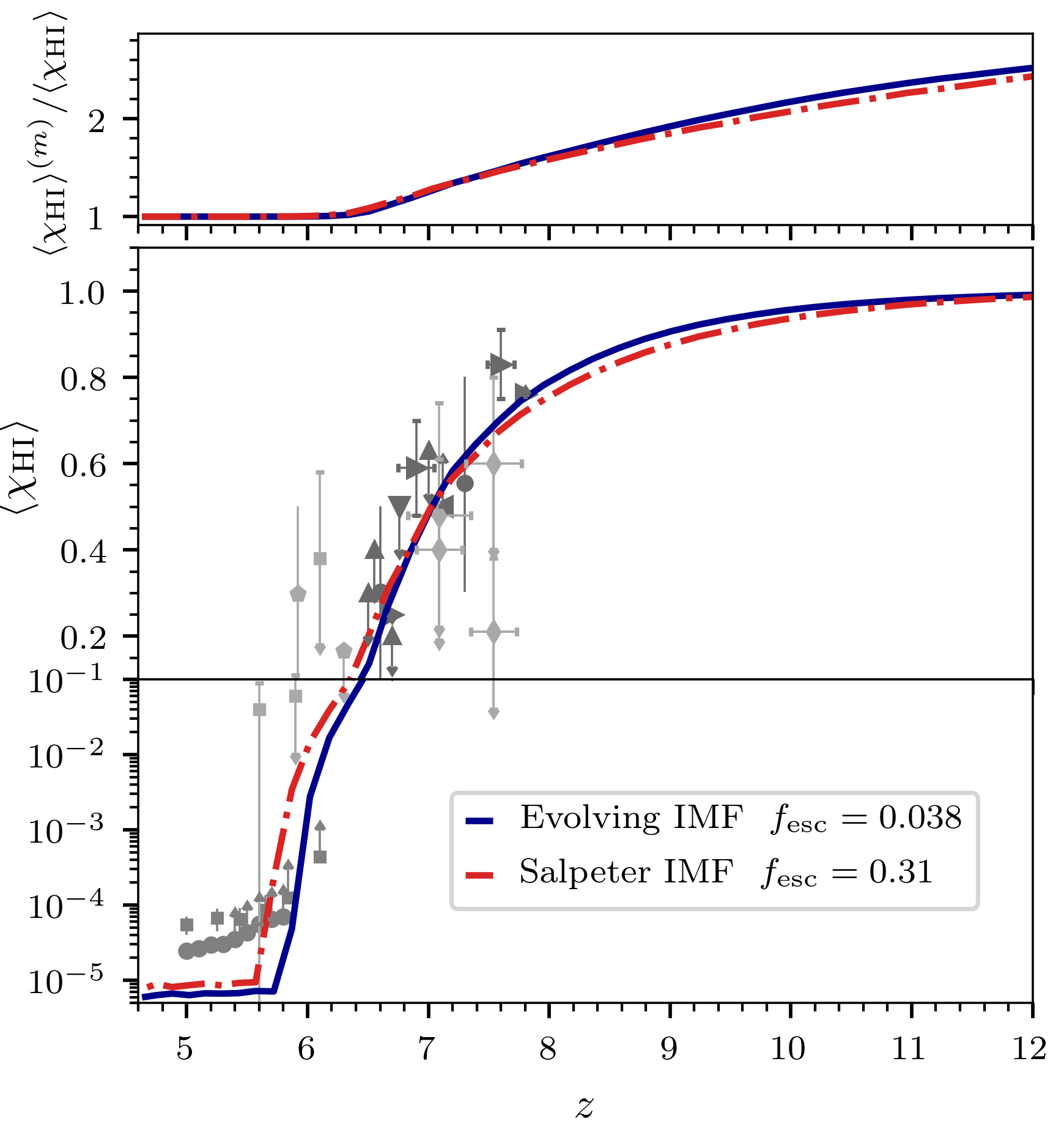}
    \caption{Reionisation history. \textit{Top:} Ratio of the mass- and volume-averaged neutral hydrogen fraction. \textit{Bottom:} Volume averaged neutral hydrogen fraction as a function of redshift. In each panel, we show results for the Salpeter IMF (dash-dotted red line) and the evolving IMF (solid dark blue line) models. Grey points indicate observational constraints from: GRB optical afterglow spectrum analyses \citep[light pentagons;][]{Totani2006, Totani2014}, quasar sightlines \citep[faint squares;][]{Fan2006}, \citep[medium bright circles;][]{Bosman2022} Lyman-$\alpha$ LFs \citep[dark upward triangles;]{Konno2018,Kashikawa2011,Ouchi2010,Ota2010,Malhotra2004}, Lyman-$\alpha$ emitter clustering \citep[dark downwards triangles;][]{Ouchi2010}, the Lyman-$\alpha$ emitting galaxy fraction \citep[dark leftwards triangles;][]{Pentericci2011, Schenker2012, Ono2012, Treu2012, Caruana2012, Caruana2014, Pentericci2014}, Lyman-$\alpha$ equivalent widths \citep[dark rightwards triangles;][]{Mason2018a, Mason2019, Bolan2022}, dark pixels \citep[light squares;][]{McGreer2015}, and damping wings \citep[light diamonds;][]{Davies2018, Greig2019}.}
    \label{fig:hist_ion}
\end{figure}

\section{Baselining the model against observed data sets}
\label{sec_baselining_model_against_observations}

In the following, we analyse and compare the simulation results for two IMF models: (i) a constant Salpeter IMF ranging from $0.1\msun$ to $100\msun$, and (ii) an evolving IMF ranging from $0.1\msun$ to $100\msun$ and becoming increasingly top-heavy towards higher redshift and lower metallicities, characterised (as described in Sect. \ref{subsec_evolving_IMF}) by shifting the transition from a Salpeter to a log-flat slope to higher star masses ($\leq100\msun$). 
For both IMF models, we adjusted the three free model parameters in {\sc astraeus}. The normalisation of the ionising escape fraction, $f_\mathrm{esc}^0$, was adjusted to fit the IGM neutral hydrogen fraction constraints from Lyman-$\alpha$ emitters, quasar absorption spectra and gamma ray bursts (GRBs), and the optical depth measured by Planck \citep{planck2018}, while the maximum star formation efficiency, $f_\star$ and the SN wind coupling efficiency, $f_w$, were adjusted to reproduce the observed dust-attenuated ultraviolet (UV) luminosity functions (LFs) at $z=5-12.5$.
We chose these eight UV LFs to fit the galaxy model parameters $f_\star$ and $f_w$. However, the more abundant observational constraints at lower redshifts ($ z\lesssim9$) effectively determine these two redshift-independent parameter values; therefore, the UV LF at $z=12.5$, which is of most interest, is not constraining the model.
We go on to detail how we found the best-fit parameters and discuss their implications.

\subsection{UV luminosity functions}
\label{subsec_UV_LFs}

For both IMF models, we derive the intrinsic and observed UV LFs from our simulations by computing the intrinsic ($L_c$) and dust-attenuated UV luminosities ($L_c^\mathrm{obs}$) for each galaxy in our simulation box, as detailed in Sect. \ref{subsec_metallicity_dust}. The simulated observed UV LFs are then compared to the observational results at $z=5-12.5$ to constrain the maximum star formation efficiency $f_\star,$ and SN feedback wind coupling efficiency $f_w$. The best-fit values for each of our two IMF model were found according to the following steps. 

At each redshift, we derive a $\chi_z^2$ characteristic, which we calculate from the linearly interpolated logarithmic number density values of the simulated and observed UV LFs and assume Poisson uncertainties on the simulated values. We then minimise the total $\chi^2$ value across all redshifts, $\chi^2_{\rm total} = \sum_z \chi^2_z$ to identify the $f_\star$ and $f_w$ best-fit values listed in Table \ref{table_model_params_fsdynamic}.\footnote{We note that we discard parameter sets that are in relative significant disagreement with the observational constraints at one or more redshifts, even if the total $\chi^2_\mathrm{total}$ is similar to the best-fit parameter sets.} Although our $\chi^2$ fitting procedure considers constraints from all redshifts, we stress that our results are biased to lower redshifts ($z\leq9$) where observational constraints are tighter.
While increasing $f_w$ suppresses the star formation in lower mass galaxies ($M_{h} \lesssim10^{10.3}\msun$ and $10^{10.4}\msun$ for the Salpeter IMF and evolving IMF model at $z=6$), thus repressing and flattening the faint end of the UV LF, increasing $f_\star$ raises effectively the star formation in more massive galaxies shifting the bright end of the UV LF to higher UV luminosities. 
Since the evolving IMF model's top-heavier IMF leads to a higher abundance of massive stars and, thus, a lower mass-to-UV light ratio, we need to reduce its star formation efficiency compared to the Salpeter IMF to fit the observed UV LFs. We identify a $2.5\times$ lower star formation efficiency of $f_\star=0.01$ and a $1.5\times$ higher SN wind coupling efficiency of $f_w=0.3$ than the Salpeter IMF model's values of $f_\star=0.025$ and $f_w=0.2$.

We show the corresponding intrinsic (dashed lines) and observed (solid lines) UV LFs at $z=6-11$, $12.5$ and $15$ for the evolving (blue lines) and Salpeter (red lines) IMF models along with observational data (grey and black points) in Fig.~\ref{fig:UVLF}. Overall, both IMF models agree within the observational uncertainties with the observations at $z=5-11$ but somewhat underestimate the abundance of $M_\mathrm{UV}<-19$ galaxies at $z=12.5$. Interestingly, despite the redshift dependency of the evolving IMF model, we find the observed UV LFs of both IMF models to be very similar across all redshifts; they consistently differ by no more than approximately $0.2-0.3$ dex within the range where we possess sufficient statistics ($\lg(N)\gtrsim-5.8$ with 10 galaxies per bin and a bin width of $0.5$ mag) and star formation rate histories (SFHs) have converged.\footnote{In \citet{Hutter2021a}, we establish the latter to be valid for $M_{h}\geq10^{8.6}\msun$, corresponding to $M_{\rm UV}\gtrsim-14$ $(-16)$ at $z=6$ $(12.5)$ for the Salpeter and evolving IMF models, denoted by the vertical black dashed and solid lines, respectively.}
One would intuitively expect that the UV LF of the evolving IMF evolves less significantly at higher redshifts than the UV LF of the Salpeter IMF, since its IMF becomes increasingly top heavy, reducing the mass-to-UV light ratio. However, the rising abundance of massive stars towards higher redshifts results in both a more efficient suppression of star formation through stronger SN feedback and a lower mass-to-light ratio, explaining why the redshift evolution of the UV LFs of the two IMF models are so similar.

From Fig.~\ref{fig:UVLF}, we can see that the UV LF reflects the hierarchical growth of galaxies. Firstly, as cosmic time progresses, galaxies grow in mass through mergers and gas accretion, moving the UV LF to higher luminosities. For instance, while the most luminous galaxies at $z=15$ have UV luminosities of $M_\mathrm{UV}\simeq-20$, the brightest at $z=6$ are about $40\times$ brighter reaching $M_\mathrm{UV}\simeq-24$.
Secondly, more low-mass galaxies form with cosmic time. These galaxies (i) exhibit bursty star formation caused by SN feedback and radiative feedback from reionisation (most effective in low-mass halos) and (ii) are likely to be consumed by mergers in the vicinity of more massive galaxies. These processes make the redshift evolution of the UV LF's faint end more complex, effectively comprising a combination of positive and negative luminosity and number density evolution, with low-mass galaxies brightening and fading as well as forming and being consumed by merging. However, overall the number density of faint galaxies with $M_\mathrm{UV}=-16$ increases by about a factor $30$ from about $10^{-3}$~Mpc$^{-3}$ at $z=15$ to $3\times10^{-2}$~Mpc$^{-3}$ at $z=6$ for both IMF models. 
Moreover, as both SN and radiative feedback affect more massive galaxies with decreasing redshift and gas-poor low-mass galaxies can attain fainter UV luminosities, the slope of the faint end of the UV LF flattens. At the same time, the bright end steepens with decreasing redshift, as the UV radiation of more massive and thus more luminous galaxies is increasingly attenuated by dust.\footnote{The increasing importance of dust attenuation results mainly from our assumption of the galaxies' dust-to-virial radius ratio to increase with rising redshifts, motivated by the gas seemingly covering a larger volume fraction with increasing redshift as ALMA [C{\small II}] observations suggest \citep{Fudamoto2022}.}

Finally, we briefly discuss our models' predictions on the evolution of the UV LFs at $z\gtrsim10$. We find the slope of the bright end ($M_{\rm UV}\lesssim -19$) to remain essentially constant, changing by no more than a few per cent, and the number density at $M_\mathrm{UV}\gtrsim-20$ to decrease by less than $1$~dex from $z=10$ to $z=12.5$. This minor redshift evolution is in rough agreement with the observational findings \citep[e.g.][]{Harikane2023a, Harikane2023b, Ferrara2023, Castellano_early_2022} and continues up to $z=15$. However, at $z=12.5$, both our IMF models slightly underpredict the abundance of $M_\mathrm{UV}<-19$ galaxies. Should spectroscopic follow-up observations validate these abundances and the prevalence of bright galaxies show only a minor decline towards even higher redshifts, as photometric detections of $z\gtrsim13$ galaxies imply \citep{Donnan2023a, Finkelstein2023, Harikane2023a}, we would need to increase our model's star formation efficiency towards higher redshifts \citep[e.g.][]{Dekel2023} or invoke contributions from efficiently accreting massive black holes \citep[e.g.][]{Pacucci2022}.

\paragraph*{} 
In summary, despite the evolving IMF becoming more top-heavy towards higher redshifts, we do not observe any alteration in the evolution of its UV LF compared to the Salpeter IMF. Any reduction in the mass-to-UV light ratio towards higher redshifts is compensated by stronger SN feedback, which leads to an increased suppression of star formation in addition to the overall assumed lower star formation efficiency.  

\begin{figure*}
    \centering
    \includegraphics[width = \textwidth]{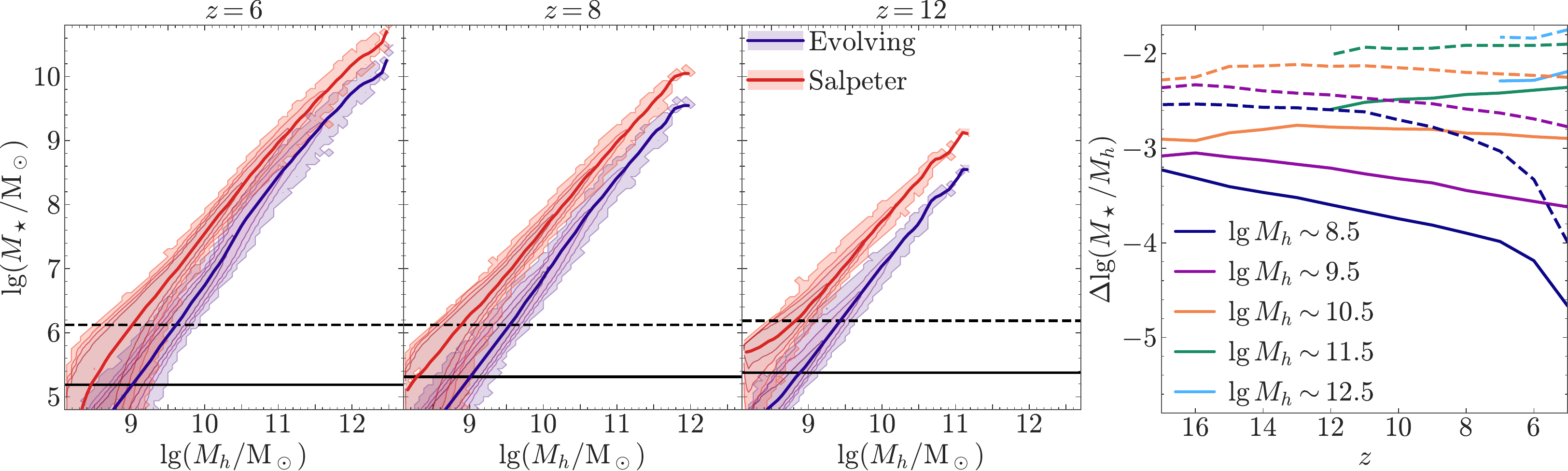}
    \caption{Redshift evolution of the stellar to halo mass relation. {\it Left:} Relation between the stellar and halo mass for the Salpeter IMF (red) and evolving IMF (blue). Coloured thick solid lines depict the medians of the distributions of stellar masses (coloured contours) at a given halo mass. Horizontal black solid and dashed lines denote the stellar mass thresholds in the evolving IMF and Salpeter IMF simulations, respectively, below which the stellar masses have not converged for all galaxies due to the mass resolution limit of the {\sc vsmdpl} simulation.
    {\it Right:} Redshift evolution of the stellar-to-halo mass ratio for different halo mass bins with a width of $1$~dex for the Salpeter IMF (dashed lines) and evolving IMF (solid lines) models. The values in the panel show the central value of each halo mass bin.}
    \label{fig:Mvir_Mstar}
\end{figure*}

\subsection{Reionisation history}
\label{subsec_reionisation_history}

For both the Salpeter IMF and the evolving IMF models, we constrained the third free model parameter, $f_\mathrm{esc}^0$, within our $f_\mathrm{esc}$ parameterisation, with the observational constraints on the volume-averaged ionisation history from quasar absorption lines, GRBs and Lyman-$\alpha$ emitters, and the electron optical depth derived from the CMB measurements with Planck \citep{planck2018}. Here, we have prioritised obtaining the most similar evolutions of the volume-averaged ionisation fraction at $z\simeq7$ for both IMF models. The corresponding values for $f_\mathrm{esc}^0$ and evolutions of the global \HI fraction $\langle\chi_\mathrm{HI}\rangle(z)$ are shown in Table \ref{table_model_params_fsdynamic} and Fig. \ref{fig:hist_ion}, respectively.  

Firstly, we find the $f_\mathrm{esc}^0$ value for the evolving IMF model ($0.038$) to be about $8\times$ lower than for the Salpeter IMF ($0.31$), decreasing the $f_\mathrm{esc}$ values for all halo masses. The evolving IMF model's decrease in $f_\mathrm{esc}$ values arises mainly from its higher ionising emissivities per stellar mass stemming from the higher abundance of massive stars generated by its top-heavier IMF.

Secondly, although the derived $f_\mathrm{esc}^0$ values lead to very similar ionisation histories for both IMF models, we find the evolving IMF model to exhibit a higher electron optical depth ($\tau=0.0580$) than the Salpeter IMF model ($\tau=0.0543$). This difference is due to the electron optical depth tracing the mass-averaged and not volume-averaged ionisation fraction. In the evolving IMF model, the spatial distribution of the ionising emissivity emerging from the underlying galaxy population follows the underlying gas density distribution more closely (see the ratio of the global mass- and volume-averaged \HI fractions in the top panel of Fig. \ref{fig:hist_ion}), shifting the ionisation history to lower redshifts at fixed electron optical depth. 

Finally, we note that both our IMF models underpredict the observational constraints on $\langle\chi_\mathrm{HI}\rangle$ at $z<5.4$ by a factor $\sim2-3$. As highlighted in \citet{Ocvirk2021}, reproducing these observational constraints requires the total ionising emissivity to drop towards lower redshifts at $4\lesssim z \lesssim6$, driven by the strong radiative suppression of star formation in low-mass halos and the escape fractions steeply declining with halo mass. While our {\sc astraeus} model includes radiative feedback from reionisation that significantly suppresses star formation in low-mass halos at $z\simeq6$ \citep[c.f. Fig.~5 in][]{Hutter2021a}, our models' too small $\langle\chi_\mathrm{HI}\rangle$ values indicate that (i) our parameterisation of $f_\mathrm{esc}$ needs to reflect an even steeper decline towards higher halo masses, or (ii) our model of radiative feedback from reionisation needs to be stronger or less delayed, or (iii) we miss the dense small-scale gas distribution. We will explore this in detail in future work.

\begin{figure*}
    \centering
    \includegraphics[width = \textwidth]{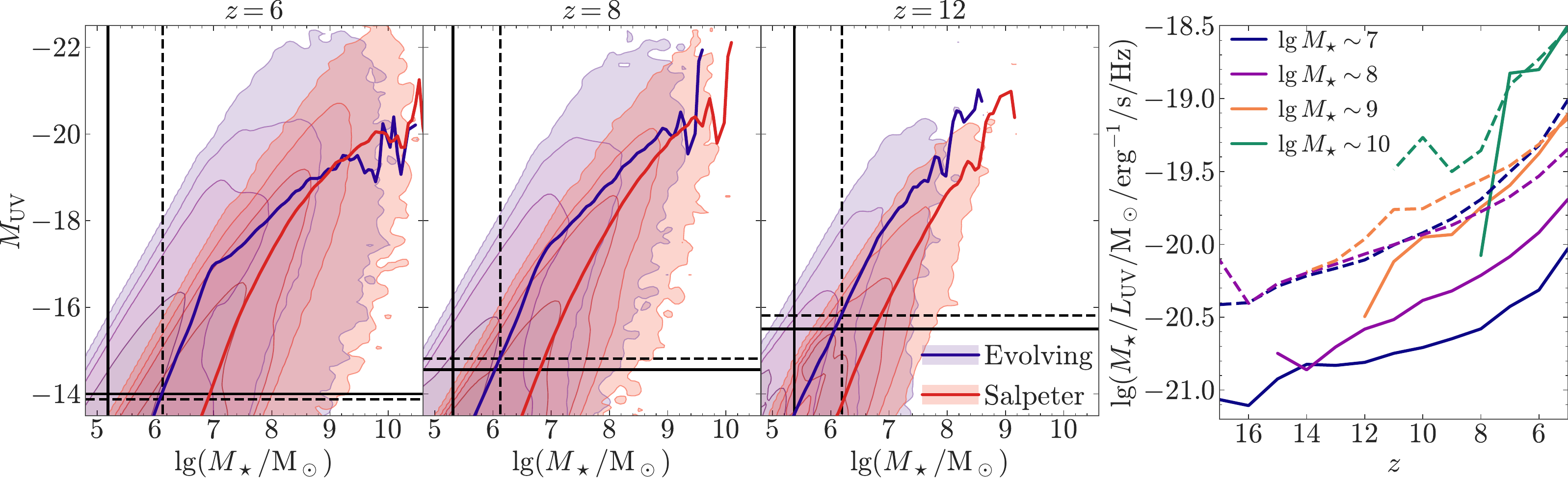}
    \caption{Redshift evolution of the observed UV luminosity to stellar mass relation. {\it Left:} Relation between the observed UV luminosity and stellar mass for the Salpeter IMF (red) and evolving IMF (blue). Coloured thick solid lines depict the medians of the distributions of observed UV luminosities (coloured contours) at a given stellar mass. Black solid and dashed lines denote the stellar mass (vertical) and UV luminosity (horizontal) thresholds in the evolving IMF and Salpeter IMF simulations, respectively, below which the stellar masses and observed UV luminosities have not converged for all galaxies due to the mass resolution limit of the {\sc vsmdpl} simulation. Median values outside the corresponding contours reflect low number statistics.
    {\it Right:} Redshift evolution of the observed UV luminosity-to-stellar mass ratio for different stellar mass bins with a width of $1$~dex for the Salpeter IMF (dashed lines) and evolving IMF (solid lines) models. The values in the panel show the central value of each stellar mass bin.}
    \label{fig:MUV_Mstar}
\end{figure*}

\section{Effects of an evolving IMF on galaxy evolution and reionisation}
\label{sec_results}

We  go on to a discussion of how going from a constant Salpeter IMF to an evolving IMF changes the relations between galaxy properties, including their star formation rates, stellar masses, gas metallicities, and dust masses. 

\subsection{Main characteristics of an evolving IMF in galaxy properties}
\label{subsec_light_to_mass_ratio}

First, to lay the foundation for understanding how all fundamental relations between galaxy properties change for an evolving IMF, we describe how such a change in the IMF alters the build-up of stellar mass (Sect.~\ref{subsubsec_stellar_to_halo_mass}) and the light-to-stellar mass ratio (Sect.~\ref{subsubsec_light_to_mass_ratio}).

\subsubsection{Shift to lower stellar masses}
\label{subsubsec_stellar_to_halo_mass}

As mentioned in Sect. \ref{subsec_UV_LFs}, we find that the evolving IMF model with its lower mass-to-light ratio requires overall lower star formation efficiencies than the Salpeter IMF model to reproduce the observed UV LFs. An important consequence of these lower star formation efficiencies is the slower buildup of stellar mass. As we can see from the relation between the stellar and halo masses in Fig. \ref{fig:Mvir_Mstar}, this results in a lower stellar-to-halo mass ratio, by $\sim0.4-1$~dex, for the evolving IMF model across all redshifts. 

The extent to which the median stellar mass is lower in the evolving IMF compared to the Salpeter IMF model for halos of the same mass exhibits two trends. Firstly, it increases towards lower halo masses, from $\sim0.5$~dex for $M_h\gtrsim10^{11}\msun$ to $\sim1$~dex for $M_h\simeq10^{9.5}\msun$ at $z=6-12$ (c.f. left panels in Fig.~\ref{fig:Mvir_Mstar}). Secondly, this difference decreases towards lower redshifts, particularly in more massive galaxies ($M_h\gtrsim10^{10}\msun$) with the stellar-to-halo mass ratio (c.f. right panel in Fig.~\ref{fig:Mvir_Mstar}) dropping by at least $0.1$~dex from $z=12$ to $5$. The first trend stems from SN feedback in the evolving IMF model becoming more efficient in suppressing star formation in galaxies located in shallower gravitational potentials (i.e. lower halo masses). This increased efficiency is due to our evolving IMF becoming more top-heavy, characterised by a higher abundance of massive stars undergoing SN events, towards lower gas metallicities and, consequently, lower halo masses. The second trend reflects that our evolving IMF becomes less top-heavy with cosmic time, resulting in less efficient SN feedback and increased star formation (despite the gravitational potentials of halos becoming shallower).

We also note that the decrease in the stellar-to-halo mass ratio with decreasing redshift for lower halo masses ($M_h\lesssim10^{10.5}\msun$) is due to gravitational potentials becoming shallower as the Universe expands, causing star formation to be SN feedback suppressed in galaxies with increasingly higher halo masses. In contrast, more massive galaxies where star formation is not SN feedback limited show a roughly constant stellar-to-halo mass ratio across all redshifts. We find the transition where a galaxy's star formation is not affected by SN feedback to occur around $M_h\sim10^{10.3}\msun$ ($10^{9.9}\msun$) and $10^{10.4}\msun$ ($10^{10.1}\msun$) for the Salpeter IMF and evolving IMF models at $z=6$ ($9$), highlighting that the larger fraction of stars exploding as SNe is counteracted by the overall lower star formation efficiency in the evolving IMF model.

The shift of stellar masses to lower values in the stellar-to-halo mass relation is one of the main features of the evolving IMF. This characteristic explains to a first-order how the relations between the main galaxy properties change when transitioning from a Salpeter IMF to an evolving IMF, which we will discuss in the following sections in detail.

\subsubsection{Reduced mass-to-light ratio}
\label{subsubsec_light_to_mass_ratio}

The second main feature of our evolving IMF is its lower galaxy-wide mass-to-light ratio compared to the constant Salpeter IMF. In Fig. \ref{fig:MUV_Mstar}, we show the observed UV luminosity-to-stellar mass relations for both IMF models at redshifts $6,8,$ and $12$ (left panels), as well as the redshift evolution of the mass-to-light ratios of galaxies at fixed stellar masses (right panel). 

Overall, galaxies with the same stellar mass are about $0-2$~mag brighter in the evolving IMF than in the Salpeter IMF model across all stellar masses and redshifts. The reason for this lies in the evolving IMF being top-heavier and thus forming a higher fraction of massive stars. As massive stars dominate the UV luminosity, galaxies of the same stellar mass contain more massive stars in the evolving IMF model and thus appear brighter.

However, we find the enhancement of the median UV luminosity at a given stellar mass to vary across stellar masses and redshifts. While the difference in the median UV luminosity at a fixed stellar mass remains around $2-3$~mag for lower mass galaxies ($M_\star\lesssim10^{7}\msun$), it decreases for more massive galaxies ($M_\star\gtrsim10^{7}\msun$) towards higher stellar masses with this decrease becoming more pronounced towards lower redshifts. For instance, at $z=12$, this difference drops from $\sim2$~mag for $M_\star=10^7\msun$ galaxies to $\sim1$~mag for $M_\star=10^{8.5}\msun$ galaxies, while it drops from $\sim3$~mag for $M_\star=10^7\msun$ galaxies to $\sim0$~mag for $M_\star=10^9\msun$ galaxies at $z=6$.
Similarly, we find the enhancement of the median mass-to-light ratio of $M_\star=10^9\msun$ galaxies to decrease from $\sim0.7$~dex at $z=12$ to $0$~dex at $z=6$ (c.f. right panel).

\begin{figure*}
    \centering
    \includegraphics[width = \textwidth]{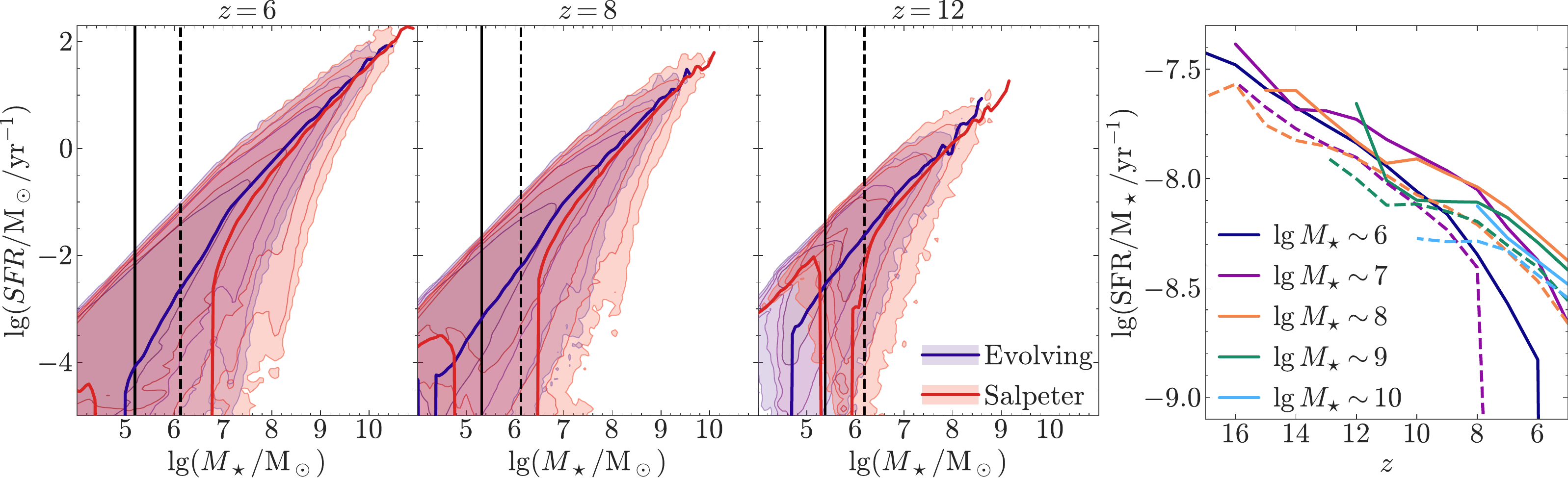}
    \caption{Redshift evolution of the SFR to stellar mass relation. {\it Left:} Relation between SFR and stellar mass for the Salpeter IMF (red) and evolving IMF (blue). Coloured thick solid lines depict the medians of the distributions of SFRs (coloured contours) at a given stellar mass. Vertical black solid and dashed lines denote the stellar mass threshold in the evolving IMF and Salpeter IMF simulations, respectively, below which the stellar masses have not converged for all galaxies due to the mass resolution limit of the {\sc vsmdpl} simulation. 
    {\it Right:} Redshift evolution of the specific SFR for different stellar mass bins with a width of $1$~dex for the Salpeter IMF (dashed lines) and evolving IMF (solid lines) models. The value in the panel shows the central value of each stellar mass bin.
    }
    \label{fig:SFR_Mstar}
\end{figure*}

These trends reflect that the evolving IMF becomes more top-heavy towards lower gas metallicities and higher redshifts (see Fig. \ref{fig:IMF}). Firstly, due to having their star formation suppressed and their gas expelled by SN feedback, lower mass galaxies exhibit lower gas metallicities across all redshifts, which causes their top-heavy IMF and thus the low mass-to-light ratio of their newly forming stars to hardly evolve with cosmic time (c.f. Fig. \ref{fig:IMF})\footnote{Despite lower mass galaxies expelling more gas in the evolving IMF model, the overall IGM metallicity hardly increases compared to the Salpeter IMF due to the evolving IMF model's lower star formation efficiency.}. This explains the evolving IMF's nearly constant negative offset in median UV luminosities to the Salpeter IMF for lower mass galaxies ($M_\star\lesssim10^7\msun$). Secondly, as galaxies become more massive, they can retain their gas and metals, increasing their gas metallicities as they accumulate stellar mass. As a consequence, their IMFs become increasingly Salpeter-like. The resulting increasing mass-to-light ratio with rising stellar mass explains why the difference in the median UV luminosities between the evolving IMF and Salpeter IMF models decreases with increasing stellar masses. This decrease becomes more pronounced towards lower redshifts because the trend of the evolving IMF approaching the Salpeter IMF becomes more distinct. The latter is due to the redshift evolution of the evolving IMF, which is the strongest for high metallicities. 

These metallicity and redshift dependencies of the evolving IMF also affect the redshift evolution of the absolute mass-to-light ratio values, as shown in the right panel of Fig. \ref{fig:MUV_Mstar}. While the mass-to-light ratio increases with decreasing redshifts across all stellar masses for both models, the evolving IMF model exhibits a steeper increase with redshift than the Salpeter model. Generally, an increase in the mass-to-light ratio towards lower redshifts is expected: galaxies accumulate stellar mass as they evolve, but their UV luminosities are dominated by the short-lived ($\lesssim30$~Myr) massive stars and thus only trace the relatively newly forming stars. Given that our evolving IMF approaches the Salpeter IMF towards lower redshifts and higher stellar masses, the abundance of UV-bright short-lived massive stars drops. This drop in the massive star abundance reduces the mass-to-light ratio of the newly forming stars and thus the overall mass-to-light ratios of galaxies as they grow with cosmic time.

Finally, it is important to note that the evolving IMF model shows a larger scatter in the UV luminosity values for a given stellar mass than the Salpeter IMF model. For instance, while the UV luminosities for galaxies with $M_\star\sim10^8\msun$ range only from $M_{\rm UV}\sim-12$ to $-19$ at $z=6$ in the Salpeter IMF model, they extend up to $M_{\rm UV}\sim-21$ in the evolving IMF model. There are two reasons for this enhanced scatter. Firstly, since the more top-heavy IMF in the evolving IMF model produces a higher abundance of massive stars, fewer stars are left over after the short-lived massive stars die, giving rise to UV-fainter galaxies between consecutive starburst events. Secondly, as the evolving IMF is sensitive to the metallicity, with metal-poor galaxies having a more top-heavy IMF than metal-rich galaxies, any scatter in the galaxies' metallicity values due to their different accretion and merger histories enhances the scatter in the UV.

\paragraph*{}
In summary, the main characteristics of an IMF that becomes increasingly top-heavy towards higher redshifts and lower metallicities are: (1) the slower build-up of stellar mass resulting in a shift to lower stellar masses in the stellar-to-halo mass relations and (2) the reduced mass-to-light ratio due to the higher abundance of massive stars.

\subsection{Star formation main sequence}
\label{subsec_starformation_main_sequence}

We next discuss how the star formation main sequence depends on the assumed IMF model. For this purpose, we show the relation between the star formation rate (SFR) and stellar mass at $z=6,8$ and $12$ and the redshift evolution of the specific star formation rate $\mathrm{sSFR}=\mathrm{SFR}/M_\star$ of galaxies with fixed stellar masses in Fig. \ref{fig:SFR_Mstar}. We derive the SFR of each galaxy in our simulations by evaluating how much stellar mass formed in the current time step, $M_\star^\mathrm{new}/\Delta t$ with $\Delta t$ being the current time step's length.

Firstly and most notably, we find that the median star formation main sequence remains unchanged for galaxies where star formation is driven by gas accretion and not SN feedback limited ($M_\star\gtrsim10^7\msun$), when transitioning from the Salpeter IMF to the evolving IMF model. This independence of the SFR -- stellar mass relation from the chosen IMF model can be explained as follows: The build-up of stellar mass depends on (i) the SFR and (ii) the dark matter assembly history. For each IMF, the SFR is essentially constrained by the UV luminosities of the observed galaxy populations; that is, for a more top-heavy IMF the lower stellar mass-to-observed UV luminosity ratio is compensated by a lower star formation efficiency to reproduce the observed UV LFs as described in Sect.~\ref{subsec_UV_LFs}. Since the build-up of dark matter and thus gas available for star formation in galaxies is self-similar (i.e. $\dot{M}_h/M_h\simeq \mathrm{const.}$) the fractional increase in halo or gas mass is essentially independent of the halo mass. This self-similarity has the consequence, that for different assumed star formation efficiencies ($f_\star)$, galaxies of similar stellar masses will have different halo masses, but similar SFRs. In scenarios where the star formation efficiency is reduced, galaxies of a given stellar mass are located in more massive halos where more gas is available for star formation, such that the stellar mass gained ($\mathrm{SFR}\propto f_\star M_\mathrm{g}$) remains the same as in a scenario with a higher star formation efficiency where galaxies of the same stellar mass reside in less massive halos.
While this picture holds for the median star formation main sequence relation, it ignores the halo-mass-dependent scatter of the DM assembly histories that cause the scatter in SFR values. As, on average, DM assembly histories become less diverse towards higher halo masses, we find that the scatter in SFR values in scenarios with a lower star formation efficiency is smaller. This dependence explains why the scatter in SFR values is smaller in the evolving IMF than in the Salpeter IMF model.

Moreover, the SFR in galaxies with lower stellar masses ($M_\star\lesssim10^7\msun$) is limited by SN feedback and therefore no longer scales with the gas accretion rate, which causes the median SFRs of our IMF models having different SN feedback efficiencies to diverge. The transition at which SN feedback becomes significant in suppressing star formation is visible for the Salpeter IMF; its median SFR at $z=6$ ($12$) starts to drop more below $M_\star\lesssim10^{7.5}\msun$ ($10^{6.5}\msun$), corresponding to $M_{h}\lesssim10^{10}\msun$ ($10^{9.5}\msun$). However, for the evolving IMF, we find this drop to be shifted to stellar masses below the convergence limit (see Appendix B in \citealt{Hutter2021a}).\footnote{This shift to lower stellar masses is also larger than what could be explained by the $\sim10\times$ smaller stellar-to-halo mass ratio, given that SN feedback depends primarily on the gravitational potential (halo mass).} Instead of dropping abruptly, the median SFR-stellar mass relation in the evolving IMF model becomes steeper towards lower redshifts at $M_\star\lesssim10^{7}\msun$. At first sight, this median SFR, which exceeds that of the Salpeter IMF model, seems to be at odds with the higher SN wind coupling efficiency $f_w$ and the higher fraction of stars turning into SN due to the more top-heavy IMF. However, a higher fraction of massive stars also implies that a higher fraction of a stellar population's SN energy is injected sooner (i.e. in the same time step) rather than being delayed (to subsequent time steps). As a result, the SFR is less driven or affected by SN feedback from earlier stellar populations (at previous time steps) compared to the Salpeter IMF model, resulting in the evolving IMF model's SFR histories being less extreme in its minimum and maximum SFR values and contributing to the smaller scatter in SFR values.

Finally, we find the median sSFR values in all stellar mass bins to decrease with decreasing redshift for both IMF models (see the right panel in Fig. \ref{fig:SFR_Mstar}), which is in agreement with previous observational and theoretical works \citep[e.g.][]{Labbe2013, Stark2013, Hutter2014}. This decrease in the sSFR values is expected because the gravitational potentials of galaxies become shallower towards lower redshifts, which again allows SN feedback to limit star formation in more massive galaxies.

\begin{figure*}
    \centering
    \includegraphics[width = \textwidth]{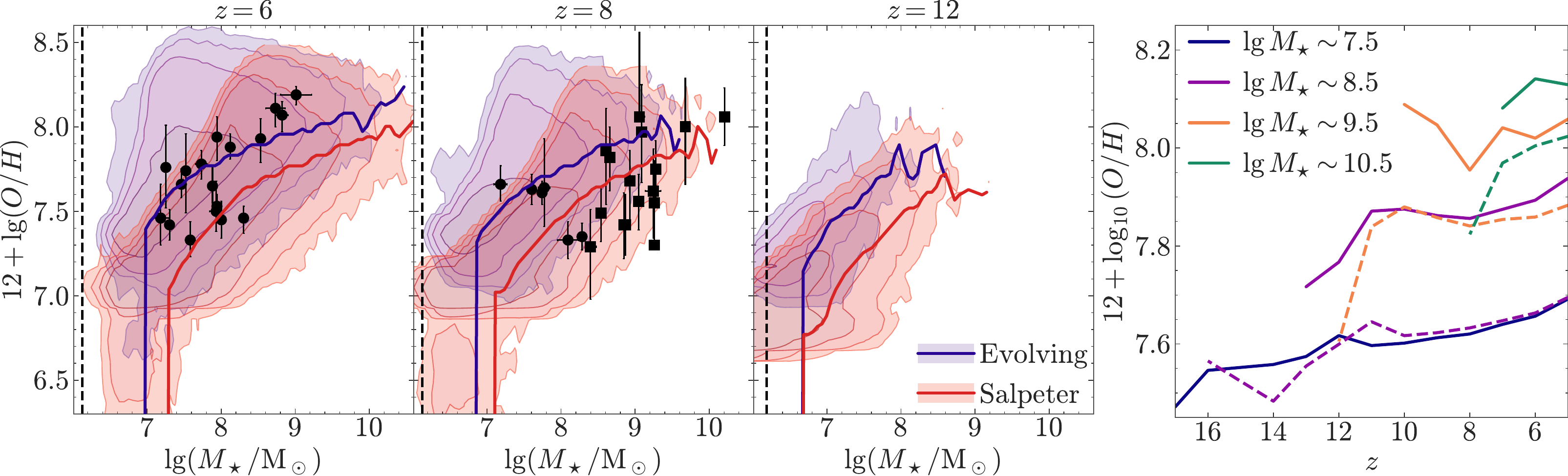}
    \caption{Redshift evolution of the oxygen-based gas metallicity to stellar mass relation. {\it Left:} Relation between the oxygen-based gas metallicity and stellar mass for the Salpeter IMF (red) and evolving IMF (blue). Coloured thick solid lines depict the medians of the distributions of gas-phase metallicities (coloured contours) at a given stellar mass. Black squares and circles show the observational results from \citealp{Heintz2023c} ($z\simeq7.1-9.5$) and \citealp{Curti2023a} ($z\simeq5.8-9.4$) converted to a Salpeter IMF stellar masses according to \citet{Speagle2014}, respectively. Observations at $z<7$ are shown in the $z=6$ panel and observations at $7\leq z<10$ in the $z=8$ panel. Vertical black dashed lines denote the stellar mass threshold in the Salpeter IMF simulations, below which the stellar masses have not converged for all galaxies due to the mass resolution limit of the {\sc vsmdpl} simulation. {\it Right:} Redshift evolution of the oxygen-based gas-phase metallicity for different stellar mass bins with a width of $1$~dex for the Salpeter IMF (dashed lines) and evolving IMF (solid lines) models. The values in the panel show the central value of each stellar mass bin.}
    \label{fig:Z_Mstar}
\end{figure*}

\subsection{Evolution of metallicity and dust mass}
\label{subsec_metallicity_dust}

During reionisation, SNe play a crucial role as primary sources of metals and thus contribute to the formation of dust. While the abundance and types of elements produced depend on the IMF of the forming stars, in this work, we focus on how the evolution of the total metal and dust content in galaxies changes as we transition from the constant Salpeter IMF to the more top-heavy evolving IMF. For this purpose, we explored how the galaxies' gas-phase metallicity and dust masses correlate with their stellar masses and UV luminosities. While the existing literature primarily discusses the stellar mass-metallicity relation, it is important to note that inferring the stellar mass from observations requires assuming a specific IMF. Therefore, we have also included how the gas-phase metallicity depends on the observed UV luminosity -- with the latter being a directly observable quantity and, thus, independent from the IMF in observational inferences. We note that such a relation will not directly probe the underlying 3D relation between stellar mass, metallicity, and SFR \citep[e.g.][]{Mannucci2010, LaraLopez2010}, as the observed UV luminosity is a proxy for dust-attenuated and not intrinsic SFR. 

\subsubsection{Fundamental metallicity relation}
\label{subsubsec_Z_Mstar}

We start by discussing how the relation between the gas-phase oxygen abundance, $12 + \lg(\mathrm{O}/\mathrm{H})$, (referred to as metallicity in the following) and stellar mass, $M_\star$, the mass-metallicity relation (MZR), evolves for both our IMF models. We show the respective relations at $z=6$, $8$ and $12$, and the redshift evolution of the median metallicity for different stellar mass bins in Fig. \ref{fig:Z_Mstar}. We note that since {\sc astraeus} follows the oxygen abundance within a galaxy's metal reservoir explicitly, we derive the metallicity directly from the oxygen, ($M_\mathrm{O}$), and gas masses, ($M_\mathrm{g}$), $\lg(\mathrm{O}/\mathrm{H}) = \lg \left( \frac{M_\mathrm{O}}{M_\mathrm{g}} \frac{m_\mathrm{H}}{m_\mathrm{O} (1-Y)} \right),$ with $m_\mathrm{H}$ and $m_\mathrm{O}$ being the masses of hydrogen and oxygen atoms and $Y$ as the helium mass fraction. We assume the solar metallicity from \citet{Asplund2009}, yielding $12 + \lg(\mathrm{O}/\mathrm{H})=8.76$ as the solar metallicity normalisation. From the left panels of Fig.~\ref{fig:Z_Mstar}, we can see that the gas in a galaxy becomes increasingly metal-enriched as the galaxy builds up its stellar mass through gas accretion-driven star formation and mergers. For instance, at $z=6$, the median oxygen-based metallicity increases from $7.8$ and $7.6$ at $M_\star=10^8\msun$ to $8.1$ and $8.0$ at $M_\star=10^{10}\msun$ for the Salpeter IMF and evolving IMF models, respectively. While each episode of star formation leads to an injection of metals into the ISM, only galaxies not limited by SN feedback can retain their metal-enriched gas. In our model, low-mass galaxies ($M_\star\lesssim10^{7}\msun$ at $z=6$) have most of their gas ejected and cannot hold onto their metal-enriched gas, causing the drop of the median metallicity at $M_\star\lesssim10^{7}\msun$ in both IMF models. However, as galaxies become more massive, the fraction of gas and metals ejected by SN feedback decreases \citep[see Fig. 3 in][]{Ucci2023}, causing the median gas-phase metallicity to rise with increasing stellar mass despite accreting more pristine gas from the IGM.

Comparing the results of the evolving IMF and Salpeter IMF models, we find that galaxies with the same stellar masses contain more metal-enriched gas in the evolving IMF model. For galaxies not limited by SN feedback ($M_\star\gtrsim10^8\msun$), the enhancement in the median metallicity is $\sim0.2$~dex and hardly varies with redshift and stellar mass and can be explained as follows: The more top-heavy IMF in the evolving IMF model leads to a higher metal enrichment rate per stellar mass formed and to a higher abundance of oxygen than in the Salpeter IMF model, both increasing the median oxygen-based metallicity. We also note that the halo mass-metallicity relations are very similar for both IMFs, possibly because we adjust the star formation efficiencies to reproduce the observed UV LFs (leading to very similar halo mass-to-UV luminosity ratios) and the enhancement in the UV luminosity per stellar mass for a more top-heavy IMF is similar to the enhancement of the corresponding metal enrichment rate per stellar mass.

Figure \ref{fig:Z_Mstar} also shows that galaxies with the same stellar mass can have a range of metallicities, with the width of the simulated galaxies' metallicity values decreasing towards larger stellar masses, i.e. from $\Delta(12 + \lg(\mathrm{O}/\mathrm{H}))\simeq1.8$ ($1.2$) at $M_\star=10^8\msun$ to $0.8$ ($0.6$) at $M_\star=10^{10}\msun$ for the evolving (Salpeter) IMF model at $z=6$. This scatter in gas-phase metallicity values is driven by these galaxies different SFR values: Galaxies with higher SFR values exhibit lower metallicity values (see Fig.~\ref{fig:HFPZ} in Appendix \ref{app_mass-metallicity-SFR}), as the associated enhanced accretion of metal-poor gas from the IGM that drives the star formation dilutes the gas and decreases the metallicity \citep[see also][]{Ucci2023}. Moreover, as galaxies become more massive, the scatter in their SFR values decreases as their DM assembly and thus star formation histories become less diverse (i.e. a massive halo will always sit in a highly overdense region while a less massive halo can be located in an overdense or underdense region where its accretion of matter is reduced or enhanced, respectively; see also discussion for the scatter in the SFR values in Sect. \ref{subsec_starformation_main_sequence}).

This dependence of the gas-phase metallicity on the SFR highlights -- as has also been noted in previous studies \citep[e.g.][]{Mannucci2010, LaraLopez2010, Ucci2023} -- that the MZR is a projection of the 3D fundamental metallicity relation (FMR), which connects the stellar mass, SFR and gas-phase metallicity. As demonstrated in previous works \citep{Hunt2012, Hunt2016a, Tortora2022}, the fundamental metallicity relation can be distilled to a fundamental plane of metallicity. We perform a linear regression analysis and find our data to lie on a plane in the four-dimensional parameter space of gas-phase metallicity, stellar mass, SFR, and redshift. We derive this high-z fundamental plane of metallicity (HFPZ) for both IMF models. While the one for the Salpeter IMF yields
\begin{eqnarray}
    12 + \lg\left(\mathrm{O}/\mathrm{H}\right) &=& (0.382\pm0.064)* \lg\left(\frac{M_\star}{\msun}\right) \\
    &-& (0.067\pm0.053) * \lg\left(\frac{\mathrm{SFR}}{\msun \mathrm{yr}^{-1}}\right) \nonumber \\
    &+& (0.003\pm0.012) * z + (4.421\pm 0.57), \nonumber
\end{eqnarray}
the HPFZ for the evolving IMF shifts to higher metallicities for any point on the star formation main sequence:
\begin{eqnarray}
    12 + \lg\left(\mathrm{O}/\mathrm{H}\right) &=& (0.683\pm0.186)* \lg\left(\frac{M_\star}{\msun}\right) \\
    &-& (0.331\pm0.138) * \lg\left(\frac{\mathrm{SFR}}{\msun \mathrm{yr}^{-1}}\right) \nonumber \\
    &+& (0.032\pm0.042) * z + (2.113\pm 1.652). \nonumber
\end{eqnarray}
We note that for both IMF models, the HFPZ depends only weakly on redshift and would also be consistent with no evolution with redshift. The reason for this weak redshift evolution lies in the star formation main sequence evolving only slightly towards higher stellar masses with decreasing redshift; the corresponding slow drop of the sSFR values and thus gas fraction $M_\mathrm{gas}/(M_\mathrm{gas}+M_\star)$ towards lower redshifts causes the slightly rising metallicities.
Finally, we briefly comment on the drop in the median metallicity for galaxies with lower stellar masses ($M_\star\lesssim10^{7.5}\msun$) across all redshifts. In the Salpeter IMF model, this drop occurs at slightly higher stellar masses compared to the evolving IMF model. For instance, at $z=6$ ($z=12$), the drop arises at $M_\star\simeq10^7\msun~(10^{6.6}\msun)$ in the Salpeter model versus $M_\star\simeq10^{7.3}\msun~(10^{7}\msun)$ in the evolving IMF model. This drop results from our treatment of gas, star formation, and SN feedback in low-mass galaxies, as detailed in Sect. \ref{sec_model}. In these galaxies, either SN events have expelled all their gas, leaving them gas-depleted, or they retain gas when their star formation remains low enough to prevent SN events from ejecting all gas. As the gravitational potentials of these galaxies become shallower over cosmic time, the halo mass at which SN feedback can expel the entire gas content rises. However, despite the evolving IMF's stronger SN feedback, its drop occurs at lower stellar masses. As explained in Sect. \ref{subsec_starformation_main_sequence}, its more top-heavy IMF results in a larger fraction of SN events occurring in the current time step, maintaining a low SFR without allowing it to oscillate as much as the Salpeter IMF model's SFR.

\paragraph{Comparison against observational data:}

We compare our stellar mass -- metallicity relations at $z\simeq6-10$ with results from JWST observations, including the JWST Advanced Extragalactic Survey (JADES) \citep{Curti2023b} as well as observations from the gravitational lensing clusters Abell 2744, RXJ-2129 and the Cosmic Evolution Early Release Science (CEERS) survey \citep{Heintz2023c}. The stellar masses inferred from these JWST observations depend on the IMF assumed when fitting the galaxy spectra, where \citet{Heintz2023c} used the Kroupa IMF and \citet{Curti2023b} adopted the Chabrier IMF. To compare this observational data with our simulated galaxy population, we convert the "observed" stellar masses to a Salpeter IMF according to \citet{Speagle2014}. At this stage, an analogous conversion to our evolving IMF is not possible, as it would require a redshift- and metallicity-dependent conversion factor obtained from corresponding spectral fitting analyses. Therefore, we only compare our Salpeter IMF model to the available data, shown as black points in the left panels in Fig.~\ref{fig:Z_Mstar}. Overall, we find that the stellar mass-metallicity combinations of nearly all observed galaxies\footnote{Outliers with higher stellar masses are likely due to our limited simulation volume.} occur also in our simulated galaxy population.

\begin{figure}
    \centering
    \includegraphics[width = 0.5\textwidth]{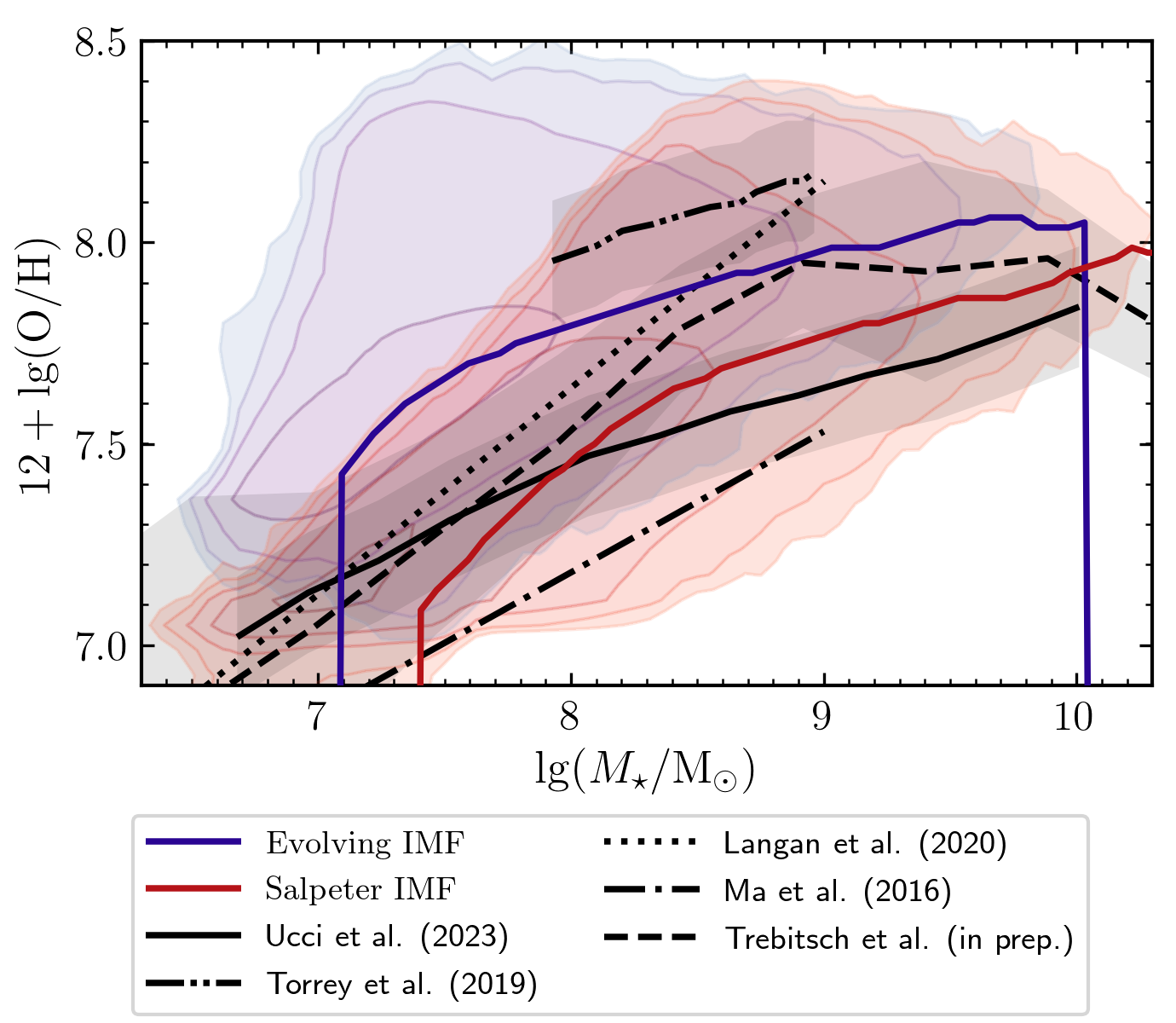}
    \caption{Comparison of the mass-metallicity relations at $z=6$ derived from different simulations. Solid blue and red lines show the median values for our evolving IMF and Salpeter IMF models. Black lines show the relations from {\sc astraeus} without dust \citep[solid with $1-\sigma$ dispersion;][]{Ucci2023}, {\sc illustris-tng} \citep[dot-dot-dashed with $1-\sigma$ dispersion;][]{Torrey2019}, {\sc firstlight} \citep[dotted;][]{Langan2020}, {\sc fire} \citep[dot-dashed;][]{Ma2016} and {\sc obelisk} (dashed with $1-\sigma$; Trebitsch et al. in prep.). We have rescaled the metallicities from all these works to our solar value of $8.76$.}
    \label{fig:Mstar_Zo_z6_compare}
\end{figure}

\paragraph{Comparison against theoretical models:}

We briefly discuss how our stellar mass-metallicity relations at $z>5$ compare to those found in other theoretical models, including the first generation of {\sc astraeus} simulations \citep{Ucci2023}, {\sc illustris tng} \citep{Torrey2019}, {\sc firstlight} \citep{Langan2020}, {\sc fire} \citep{Ma2016}, and {\sc obelisk} \citep{Trebitsch2021}.
Comparing our Salpeter IMF model results with those of the first generation of {\sc astraeus} simulations \citep{Ucci2023}, we found that the MZR has shifted by about $0.1$~dex towards higher metallicities for $M_\star\gtrsim10^{8}\msun$. This shift is due to the $2.5\times$ higher star formation efficiency, which was necessary to compensate for the dust attenuation of UV light at the UV LF's bright end that was not accounted for in \citet{Ucci2023}.

{\sc illustris tng} and {\sc obelisk} serve as the most comparable simulations in terms of metal enrichment modelling (metal production from SNe and AGB stars) and modelled galaxy masses ($M_\star\simeq10^{6-10}\msun$ and $M_\star\simeq10^{8-9}\msun$), with solar normalisations corresponding to $12+\lg(\mathrm{O}/\mathrm{H})=8.6$ and $8.69$. 
Interestingly, the {\sc illustris tng}'s MZR aligns more with our evolving IMF's MZR at $z=6$, possibly a consequence of the {\sc illustris tng} simulations assuming: (i) a Chabrier IMF, resulting in a higher abundance of massive stars per solar mass formed than our Salpeter IMF (and thus a shift of $\sim0.2$~dex to lower stellar mass) and (ii) a lower metal loading of their winds ($0.4~Z_\mathrm{ISM}$). The smaller scatter in metallicity values at fixed stellar masses in {\sc illustris tng} is likely a consequence of its smaller simulation box of $100h^{-1}$cMpc, which does not cover more extreme mass assembly histories in over-dense or under-dense regions. Extrapolations of {\sc illustris tng} results suggest a more pronounced evolution of the MZR (increase by $0.25$~dex from $z=10$ to $6$) due to the galaxies' gas fraction $M_\mathrm{gas}/M_\star$ rising toward higher redshifts. However, similarly to the findings in \citet{Ucci2023}, our simulations show no significant redshift evolution of the gas fraction, and thus, our MZR evolves very weakly with redshift.

{\sc obelisk}'s MZR, derived from the halo-averaged gas-phase metallicities (excluding dust) and stellar masses, aligns the closest with our Salpeter IMF model's MZR compared to other non-{\sc astraeus} simulations. It exceeds our Salpeter IMF model's MZR only by $0.1-0.2$~dex, likely due to assuming a Kroupa IMF, and exhibits a comparable slope at lower stellar masses ($M_\star\lesssim10^{8.5}\msun$). Notably, its MZR also shows a change in slope, although at slightly larger stellar masses.

The {\sc fire} and {\sc firstlight} simulations are both zoom-in simulations that track galaxies with stellar masses ranging from $M_\star=10^{3-9}\msun$ for {\sc fire} and $M_\star=10^{6-9.5}\msun$ for {\sc firstlight}. Both simulations adopt a Kroupa IMF, SNII yields from \citet{Woosley_Weaver1995}, and a solar metallicity normalisation of $12+\lg(\mathrm{O}/\mathrm{H})=8.9$. Despite their different implementations of physics and simulation methods, we find the {\sc firstlight}'s MZR to surpass our Salpeter IMF model's MZR by only $\lesssim0.2$~dex at $z=6-8$ for $M_\star\sim10^{7.5-8.5}\msun$. Meanwhile, the MZR from the {\sc fire} simulation lies approximately $0.3$~dex below our Salpeter IMF results. Notably, {\sc firstlight} shows a comparable redshift evolution and slope of the MZR, while the predictions from {\sc fire} suggest a more pronounced redshift evolution of the MZR, but only based on extrapolations beyond $z=6$.

\begin{figure*}
    \centering
    \includegraphics[width = \textwidth]{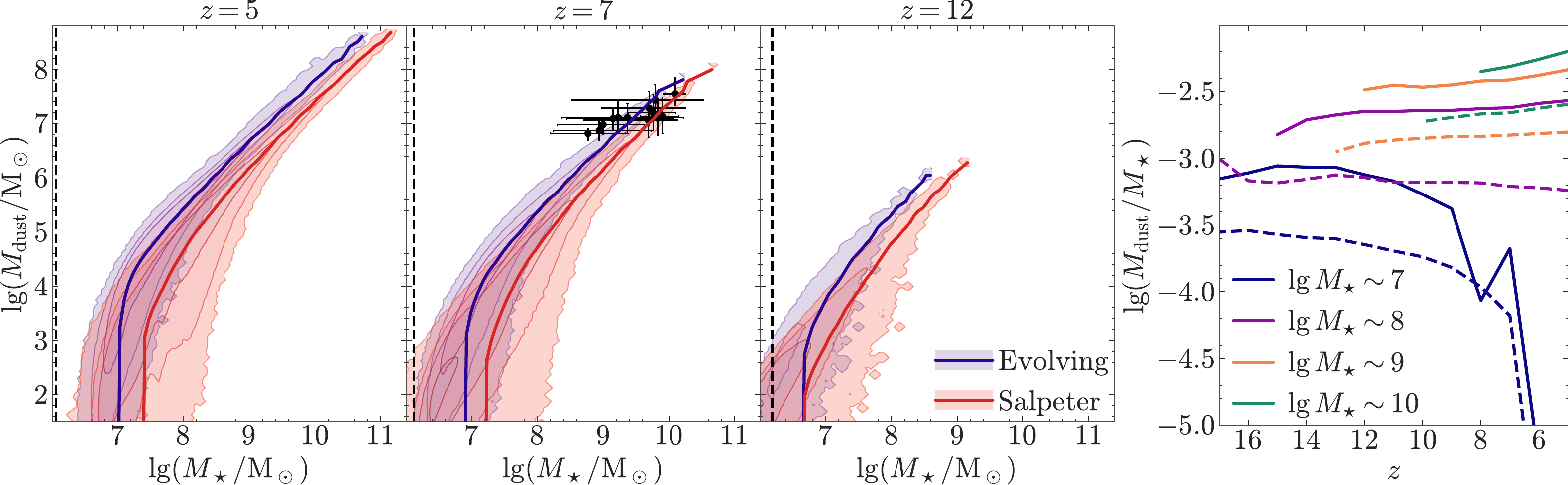}
    \caption{Redshift evolution of the dust to stellar mass relation. {\it Left:} Relation between the dust and stellar mass for the Salpeter IMF (red) and evolving IMF (blue). Coloured thick solid lines depict the medians of the distributions of dust masses (coloured contours) at a given stellar mass. Black circles show the observational results at $z\simeq7$ from the REBELS survey shown in \citealp{Dayal2022} assuming a Salpeter IMF. Vertical black dashed lines denote the stellar mass threshold in and Salpeter IMF simulations, below which the stellar masses have not converged for all galaxies due to the mass resolution limit of the {\sc vsmdpl} simulation. 
    {\it Right:} Redshift evolution of the dust-to-stellar mass ratio for different stellar mass bins with a width of $1$~dex. The values in the panel shows the central value of each stellar mass bin. Circles show median values of the REBELS galaxies.}
    \label{fig:Mdust_Mstar}
\end{figure*}

\subsubsection{The stellar mass to dust mass relation}
\label{subsubsec_dust_Mstar}

We now discuss how the relation between the dust and stellar mass alters for an evolving IMF. For this purpose, we show the respective relations for the Salpeter and evolving IMF models at $z=5$, $7$ and $12$ and the redshift evolution of the dust-to-stellar mass ratio for galaxies with fixed stellar masses in the left and right panels of Fig. \ref{fig:Mdust_Mstar}, respectively. 

Firstly, since the majority of dust is also produced in SN events, the dust mass evolves with the metal mass. Therefore, for both IMF models, the median dust mass rises with increasing stellar mass and drops at the same stellar masses towards lower stellar masses, analogous to the increasing trend of the median gas-phase metallicity with stellar mass. 

Secondly, however, we can see that the dust masses in the evolving IMF model are $\sim0.3-0.5$~dex higher than in the Salpeter IMF model for $M_\star\gtrsim10^{7-7.5}\msun$ at $z=12-6$. This shift to higher dust masses is not solely due to the $\sim0.5-1$~dex lower stellar-to-halo mass ratio discussed in Sect. \ref{subsubsec_stellar_to_halo_mass}. Unlike metals, dust -- which we consider to be a subreservoir of metal mass -- is also destroyed by SN shock waves. Therefore, due to the more top-heavy IMF, the evolving IMF model yields about $0.3$~dex lower dust-to-metal mass ratio and, thus, lower dust masses for a given halo mass than in the Salpeter IMF model. This explains (i) why the evolving IMF model's offset of the dust masses from those in the Salpeter IMF model is smaller than the shift towards lower stellar masses and (ii) why the evolving IMF model's difference between the intrinsic and dust attenuated UV LFs shown in Fig.~\ref{fig:UVLF} are smaller.

Thirdly, for galaxies with $M_\star\gtrsim10^8\msun$, the dust-to-stellar mass ratio hardly evolves with redshift, for the same reason that the median metallicity barely evolves. For less massive galaxies ($M_\star\lesssim10^8\msun$), the dust-to-stellar mass ratio decreases towards lower redshifts, as the gravitational potentials of galaxies of similar halo masses become shallower towards lower redshifts and, together with the gas, dust is more easily ejected by SN-driven winds.

Finally, we also compare our dust and stellar masses to the observational results from the ALMA REBELS survey \citep[see gray points in Fig. \ref{fig:Mdust_Mstar} from][]{Dayal2022}. Both our IMF models agree with the REBELS data within the observational uncertainties. However, it should be noted that the stellar and dust masses derived from the REBELS galaxies assume a Salpeter IMF. 

\begin{figure*}
    \centering
    \includegraphics[width = \textwidth]{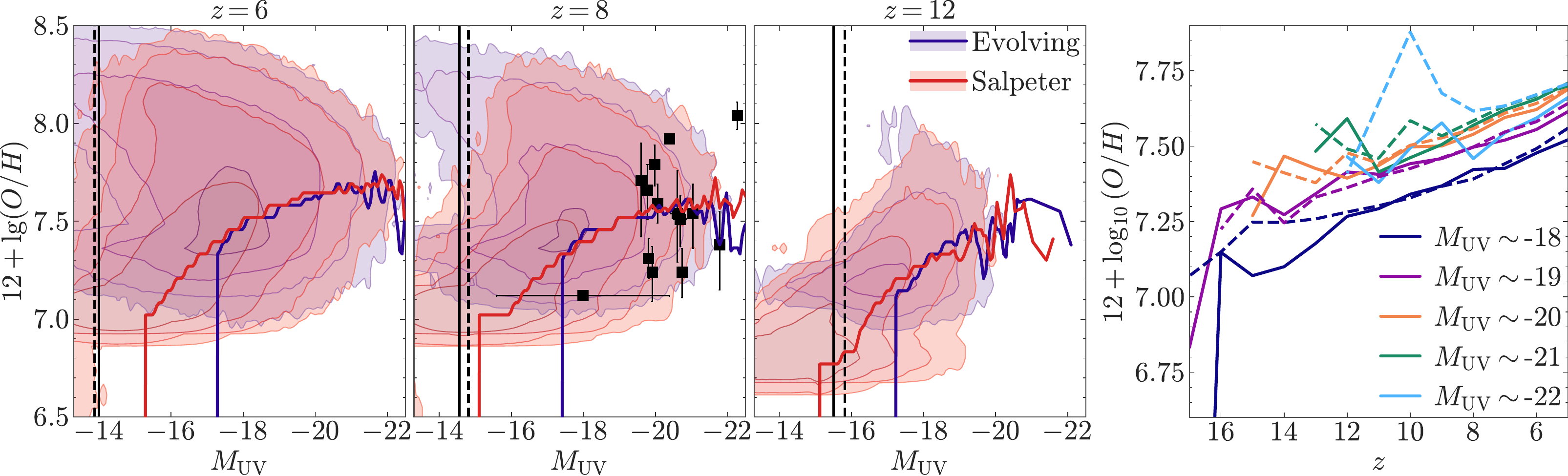}
    \caption{Redshift evolution of the oxygen-based metallicity to observed UV luminosity relation. {\it Left:} Relation between the oxygen-based gas metallicity and the observed UV luminosity for the Salpeter IMF (red) and evolving IMF (blue). Coloured thick solid lines depict the medians of the distributions of gas metallicities (coloured contours) at a given UV luminosity. Black squares show the observational results from \citet{Heintz2023c} at $z\simeq7.1-9.5$. Vertical black solid and dashed lines denote the UV luminosity threshold in the evolving IMF and Salpeter IMF simulations, respectively, below which the UV luminosities have not converged for all galaxies due to the mass resolution limit of the {\sc vsmdpl} simulation. 
    {\it Right:} Redshift evolution of the oxygen-based gas metallicity for different UV luminosity bins with a width of $1$~dex for the Salpeter IMF (dashed lines) and evolving IMF (solid lines) models. The values in the panel show the central value of each UV luminosity bin. }
    \label{fig:Z_MUV}
\end{figure*}

\begin{figure*}
    \centering
    \includegraphics[width = \textwidth]{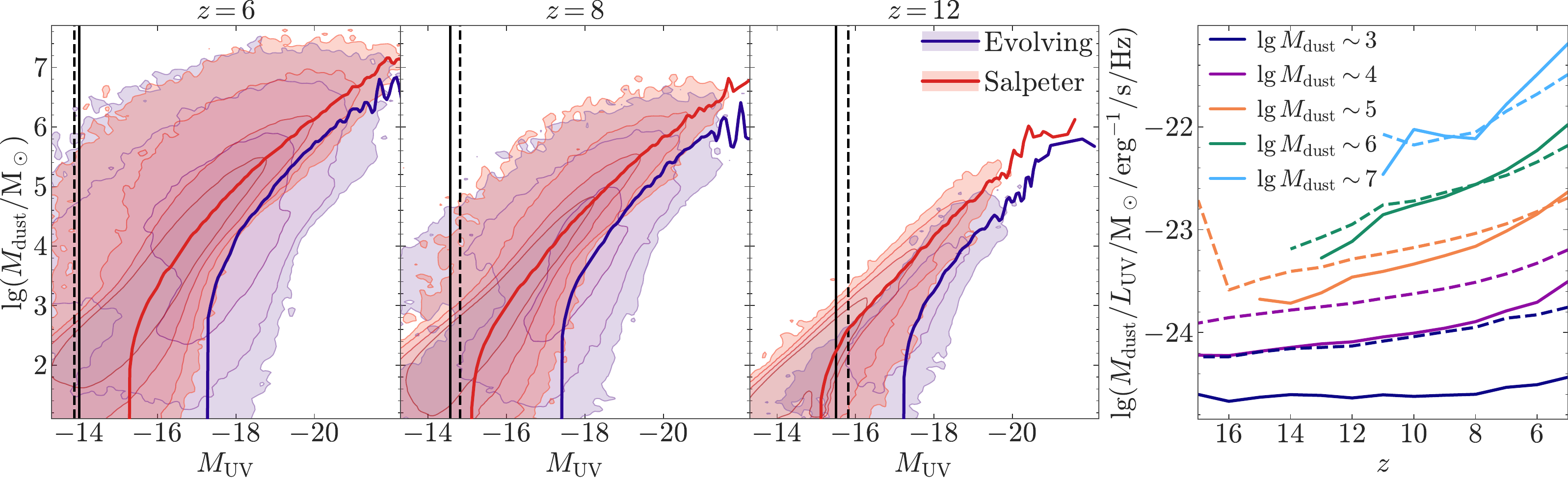}
    \caption{Redshift evolution of the dust mass to observed UV luminosity relation. {\it Left:} Relation between the dust mass and the UV luminosity for the Salpeter IMF (red) and evolving IMF (blue). Coloured thick solid lines depict the medians of the distributions of dust masses (coloured contours) at a given UV luminosity. Vertical black solid and dashed lines denote the UV luminosity threshold in the evolving IMF and Salpeter IMF simulations, respectively, below which the UV luminosities have not converged for all galaxies due to the mass resolution limit of the {\sc vsmdpl} simulation. 
    {\it Right:} Redshift evolution of the dust mass-to-UV luminosity ratio for different dust mass bins with a width of $1$~dex for the Salpeter IMF (dashed lines) and evolving IMF (solid lines) models. The values in the panel show the central value of each dust mass bin.}
    \label{fig:Mdust_MUV}
\end{figure*}

\subsubsection{Metallicity and dust mass relations with UV luminosity}

Finally, we explore how the gas-phase metallicities and dust masses of galaxies are related to the galaxies' observed UV luminosities. Our goal is to assess to which degree these relations differ between our two IMF models, as these relations are mostly independent of the IMF assumed in observational analyses, thereby providing a potential means to distinguish between different IMFs. To this end, we show the corresponding relations at $z=6,8,12$ and the redshift evolutions of the gas-phase metallicity (in $M_\mathrm{UV}$ bins with a width of $1$~dex) and dust mass-to-UV luminosity ratio (in dust mass bins of $1$~dex) in Figs.~\ref{fig:Z_MUV} and~\ref{fig:Mdust_MUV}, respectively.

Firstly, in contrast to the metallicity and dust mass relations with stellar mass, we find that the metallicities and dust masses of galaxies with a given observed UV luminosity are lower in the evolving IMF than in the Salpeter IMF model. This switch is due to the evolving IMF model's about $1$~dex lower stellar mass-to-light ratio. 

Secondly, brighter galaxies ($M_\mathrm{UV}\lesssim-17$) exhibit similar median metallicities in both IMF models, with the median value in the evolving IMF model being slightly lower, by less than $0.1-0.2$~dex, than in the Salpeter IMF model. This similarity results from the comparable halo mass-to-light ratio and halo mass -- metallicity relations discussed in Sect. \ref{subsubsec_Z_Mstar}. However, at the same time, similar bright galaxies have lower dust masses, by about $0.4-1$~dex, in the evolving IMF than in the Salpeter IMF model, a consequence of the lower dust-to-metal mass ratio in the evolving IMF model (see Sect. \ref{subsubsec_dust_Mstar}).

Thirdly, towards lower UV luminosities ($M_\mathrm{UV}\gtrsim-17$), both the median metallicity and dust mass sharply decrease in both IMF models, albeit at higher UV luminosities in the evolving IMF model ($M_{\rm UV}\simeq-17$) than in the Salpeter IMF model ($M_{\rm UV}\simeq -15$). The shift of the location of the sharp decrease to higher UV luminosities in the evolving IMF model is the combined result of the model's higher SN rate and more efficient SN-wind coupling leading to the ejection of gas and dust in more massive halos, as well as the similar halo-mass -- observed UV luminosity relationships of the IMF models. 

Then, from the right panel in Fig.~\ref{fig:Z_MUV}, we see that the median metallicities of galaxies with a given observed UV luminosity increase slightly towards lower redshifts, by about $0.3$~dex from $z=12$ to $5$ for all luminosities. This increase is primarily due to galaxies becoming more dust attenuated with decreasing redshift (c.f. intrinsic and observed UV LFs in Fig. \ref{fig:UVLF}). It is slightly steeper for the bright and thus more metal-enriched galaxies ($M_\mathrm{UV}\lesssim-20$) in the evolving IMF model, since the evolving IMF approaches the Salpeter IMF towards lower redshifts and higher metallicities.

Next, we find similar trends for the median dust mass-to-light ratios of galaxies with fixed dust masses (see right panel in Fig.~\ref{fig:Mdust_MUV}), albeit the evolving IMF model's increase with decreasing redshift is more prominent. The dust mass-to-light ratio follows essentially the stellar mass-to-light ratio. However, due to the higher rate of dust mass production per stellar mass formed in the evolving IMF model, the relations for the two IMF models are more similar, within about $0.3$~dex for galaxies with $M_{\rm dust}\gtrsim10^{5}\msun$.

Finally, we compare our model predictions for the observed UV luminosity -- metallicity relation at $z=8$ to the observed galaxy sample analysed in \citet{Heintz2023c} (black points in Fig.~\ref{fig:Z_MUV}). We note that the inferred UV luminosities of these galaxies are not corrected for magnification; we disregard galaxies with magnifications of $>2.5$ (one galaxy) and where the limited wavelength coverage prohibits the inferral of the UV. These observations agree with both IMF models, and given the similarity of our IMF models' medians and scatter, the UV luminosity -- metallicity relation does not provide constraints on the IMF.

\subsection{Ionisation topology}
\label{subsec_ionisation topology}

\begin{figure*}
    \centering
    \includegraphics[width =0.9\textwidth]{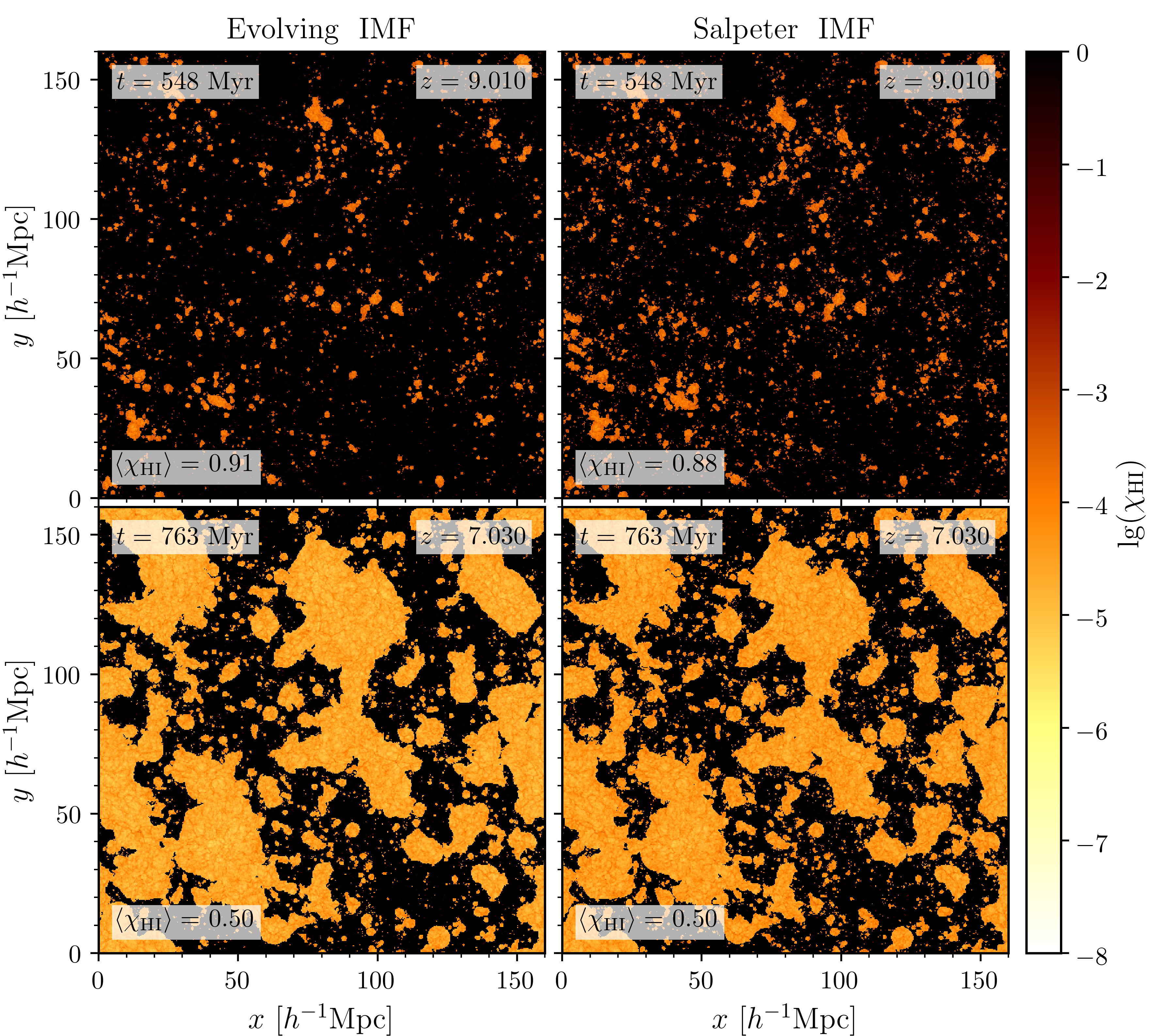}
    \caption{Neutral hydrogen fraction fields at $z\simeq9.0$ (top) and $z\simeq7.0$ (bottom) for the evolving IMF (left) and Salpeter IMF (right) models. The volume-averaged value of the neutral fraction in each cell is marked in the panels. For each redshift and model, we show a slice through the centre of the simulation box.}
    \label{fig:XHImap}
\end{figure*}

\begin{figure*}
    \centering
    \includegraphics[width = \textwidth]{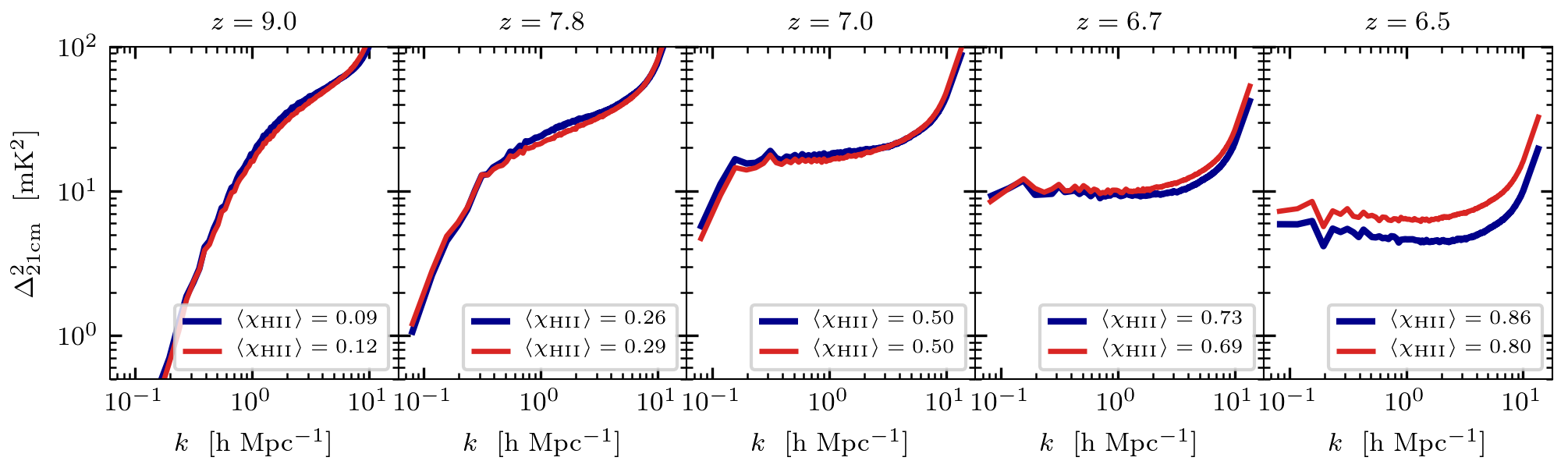}
    \caption{21cm power spectra at fixed redshifts $z$  using the best-fit parameters noted in Table \ref{table_model_params_fsdynamic}. In each panel, we show results for our two IMF models studied in this work: evolving IMF (dark blue solid line) and Salpeter IMF (red solid line). Each panel shows the average volume-averaged \HI fraction in each model at that redshift.}
    \label{fig:ps21cm}
\end{figure*}

Large radio interferometers, such as the Square Kilometre Array (SKA) or Hydrogen Epoch of Reionisation Array (HERA), will measure the 21cm signal from the intergalactic \HI gas, providing maps of the evolving spatial distribution of the \HI densities and thus of the ionised regions around galaxies during reionisation. A key statistic quantifying the size distribution of ionised regions and thus constraining the ionising emissivity distribution of the galaxy population is the 21cm power spectrum. In this section, we discuss how assuming an evolving IMF affects the topology of the ionised regions around the galaxies and the 21cm power spectrum during reionisation. For this purpose, we show the ionisation maps of both IMF models at $z\simeq9$ and $z\simeq7.4$ in Fig. \ref{fig:XHImap} and the power spectra of the simulated 21cm signal \citep[see][for the derivation of the 21cm signal from the simulated density and ionisation fields]{Hutter2020} in Fig. \ref{fig:ps21cm}.

To understand how and why the ionisation topologies differ between our two IMF models, we first analyse the dependence of the number of ionising photons escaping from a galaxy, the escaping ionising emissivity, on its halo mass. As discussed in Sects. \ref{subsec_reionisation_history} and \ref{subsec_light_to_mass_ratio}, a more top-heavy IMF leads to a higher abundance of massive stars and lower mass-to-light ratio in the UV; thus, the average intrinsic ionising emissivity per stellar mass ($\dot{Q}/M_\star$) is by about a factor $\sim7$ higher in the evolving IMF than in the Salpeter model. While compensating the higher intrinsic ionising emissivities in the evolving IMF model with a lower $f_\mathrm{esc}^0$ value leads to a similar ionisation history compared to the Salpeter IMF model (see Fig.~\ref{fig:hist_ion}), the median relation between the escaping ionising emissivities ($\dot{N}_\mathrm{ion}$) and halo mass is flatter in the evolving IMF model, i.e. more low-mass (massive) galaxies have larger (lower) intrinsic ionising emissivity values (see Fig.~\ref{fig:ionising_emissivities} in Appendix \ref{app_ionising_emissivities}). This altered distribution of the escaping ionising emissivities stems from (i) the flatter distribution of star formation rates and thus intrinsic ionising emissivities ($\dot{Q}$) for low-mass sources ($M_h\lesssim10^{10}\msun$, see Fig. \ref{fig:SFR_Mstar}) and (ii) the lower $f_\mathrm{esc}$ values for massive galaxies ($M_h\gtrsim10^{10}\msun$). However, the maximum escaping ionising emissivity values for low-mass galaxies ($M_h\lesssim10^{9.5}\msun$) are higher for the Salpeter than the evolving IMF model, as the SFR reaches higher values at similar halo masses (c.f. Fig. \ref{fig:SFR_Mstar} and accounting for the shift of $\sim1$~dex of the evolving IMF model to lower stellar masses). Together, these trends lead to the following differences in the ionisation topology. 

Firstly, the higher $\dot{N}_\mathrm{ion}$ values around $10^{10}\msun$ in the evolving IMF model lead to a higher abundance of medium-sized ionised regions compared to those in the Salpeter model. While this size difference is barely visible in the ionisation maps (Fig. \ref{fig:XHImap}), we can see an excess power around $k\simeq0.3-3h~$Mpc$^{-1}$ corresponding to spatial scales of $\sim2-20h^{-1}$~Mpc in the 21cm power spectra at $z\simeq9-7$ (Fig. \ref{fig:ps21cm}). At lower redshifts, the differences in the 21cm power spectra between our two IMF models arise from the models' diverging ionisation histories, with the evolving IMF showing a more accelerated ionisation of the IGM and, thus, lower 21cm power spectra amplitudes.

Secondly, the higher maximum $\dot{N}_\mathrm{ion}$ values for low-mass galaxies in the Salpeter model give rise to more small-sized ($\lesssim1~h^{-1}$Mpc) ionised regions than in the evolving IMF model. These regions can be seen in the predominantly neutral regions in the ionisation maps. They also cause the 21cm power spectra of the Salpeter IMF model to slightly exceed that of the evolving IMF model at small scales of $k\lesssim6~h$Mpc$^{-1}$ at $\langle\chi_\mathrm{HI}\rangle\gtrsim0.5$. 

In summary, once we adjust for the $f_\mathrm{esc}$ values in the evolving IMF model to reproduce similar ionisation histories, we find the ionisation topologies to hardly differ with minor differences occurring on medium scales of $3-30$~Mpc. Given that the median relation between the escaping ionising emissivity, $\dot{N}_\mathrm{ion}$, and halo mass is flatter in the evolving IMF than in the Salpeter IMF model, an evolving IMF results in an ionising emissivity distribution that is stronger correlated to the underlying density distribution, leading to a higher electron optical depth for similar ionisation histories.

\section{Conclusions}
\label{sec_conclusions}

We have developed the first framework that simulates the evolution of galaxies and the ionisation of the IGM and accounts for an evolving IMF in all relevant physical processes. The IMF considered here parameterises the results found in \citet{Chon2022}. It is composed of a Salpeter IMF and a log-flat IMF at the massive end, with the latter increasing towards lower gas metallicities and higher redshifts, reflecting the increasing gas temperature in galaxies due to less efficient cooling and a higher CMB temperature respectively. We have integrated this IMF into our {\sc astraeus} framework \citep{Hutter2021a, Ucci2023, Hutter2023a, Trebitsch2023}, where it shapes all IMF-sensitive processes: SN feedback, metal enrichment, and the emission of \HI ionising and UV photons. 
With this updated simulation framework, we ran two simulations: one with a non-evolving, constant Salpeter IMF and the other one assuming our evolving IMF model. The free model parameters were calibrated to match the observed UV luminosity functions at $z=5-12.5$ and the observational constraints on the global \HI fraction, $\langle\chi_\mathrm{HI}\rangle$ of the IGM from quasar sightlines, and Lyman-$\alpha$ emission from galaxies and GRBs. We analysed these two IMF scenarios to explore how an evolving IMF alters the properties of galaxies, their evolution through cosmic time, and the topology of the evolving spatial distribution of the ionised regions during reionisation. Our main results are as follows.

\begin{enumerate}
    \item The evolving IMF scenario hardly changes the redshift evolution of the UV LFs compared to the Salpeter IMF scenario. Any reduction in the mass-to-light ratio due to a more top heavy IMF towards higher redshifts is compensated by stronger SN feedback, which leads to an increased suppression of star formation in addition to the overall assumed lower star formation efficiency.  
    \item Due to the more top-heavy IMF and thus stronger SN feedback, stellar mass accumulates slower in the evolving IMF than in the Salpeter IMF model; the star formation efficiency is about $2.5\times$ lower. This results in a drop of the stellar-to-halo-mass ratio of about $0.5-1$~dex throughout reionisation, shifting the stellar mass functions by $\sim0.5-1$~dex to lower stellar masses at $z=6-12$. 
    \item As a consequence of the altered stellar-to-halo-mass relation, the evolving IMF model produces (i) UV brighter galaxies when comparing galaxies with the same stellar masses and (ii) together with its stronger and more immediate SN feedback, a broader range of UV luminosities in galaxies with similar stellar masses. Moreover, the decrease of the IMF's top-heaviness in sufficiently metal-enriched and thus massive galaxies ($M_h\gtrsim10^{8}\msun$) with increasing cosmic time causes the evolving IMF's mass-to-light ratio to approach that of the Salpeter IMF.
    \item The median star formation sequence for galaxies where star formation is driven by gas accretion ($M_h\gtrsim10^{10}\msun$, $M_\star\gtrsim10^{7}\msun$) does not change as we go from the constant Salpeter IMF to the evolving IMF. This robustness results from the self-similar growth of structures where the fractional mass growth of a DM halo is, on average, independent of its mass. For lower-mass galaxies ($M_\star\lesssim10^7\msun$), however, SFR rate values depend on the interplay between the efficiency of SN feedback and gravitational potential. Despite having more efficient SN feedback, our evolving IMF model shows fewer low SFR values for these galaxies due to them living in more massive halos with less diverse merger histories.
    \item The halo mass to gas-phase metallicity relation remains unchanged as we transition from the constant Salpeter to the evolving IMF. The independence of this relation on the IMF is attributed to calibrating star formation and SN feedback efficiencies to match the observed UV luminosity functions in both models. In contrast, the stellar mass to metallicity relation in the evolving IMF shifts towards higher metallicities compared to the Salpeter IMF, primarily due to the slower accumulation of stellar mass (partially counteracted by the increased oxygen production due to the explosion of more massive stars). Nevertheless, the lower mass-to-light ratio in the evolving IMF model causes its UV luminosity-metallicity relation to shift to metallicity values comparable to or below those of the Salpeter IMF.
    \item The dust mass assembly closely follows that of the metal mass. Yet, it is also subject to destruction by SN shock waves, leading to a lower dust-to-metal mass ratio in the evolving IMF model and galaxies where SN feedback does not suppress star formation ($M_h\gtrsim10^{10}\msun$). This lower ratio leads to a more distinct difference in the UV luminosity-dust mass relation between the two IMF models, with the evolving IMF being shifted towards lower dust masses.
    \item The lower mass-to-light ratio in the evolving IMF model results in higher production of ionising photons within galaxies than the Salpeter IMF model. Therefore, when reproducing similar reionisation histories, the average escape fraction $f_\mathrm{esc}$ in the evolving IMF is $\sim8\times$ lower.
    \item For similar reionisation histories, $\langle\chi_\mathrm{HI}\rangle$(z), the topology of the ionised regions during reionisation changes only slightly when assuming the evolving IMF compared to the constant Salpeter IMF model. During the first half of reionisation, the evolving IMF model leads to slightly more medium-sized ionised regions ($3-30$~Mpc) due to an enhancement of the escaping ionising emissivity of medium-massive galaxies ($M_h\simeq10^{10}\msun$) when the escape fraction $f_\mathrm{esc}$ follows the fraction of gas ejected by SN explosions. 
\end{enumerate}

The results from our modelling of an evolving IMF within galaxy evolution have important implications.
Firstly, our findings indicate that an IMF that becomes increasingly top-heavy towards higher redshifts does not increase the abundance of bright galaxies at $z\gtrsim10$ compared to a constant IMF. This lack of discrepancy arises because the lower mass-to-light ratio driven by the greater abundance of massive stars in a top-heavier IMF is counteracted by stronger stellar feedback, which more efficiently suppresses star formation. While we acknowledge the necessity of exploring other plausible IMF parameterisations, such as the one assumed in \citet{Steinhardt2022a}, to confirm this self-regulation and ensure the robustness of the star formation main sequence and the metallicity to halo mass relation against different IMF models, we also emphasise that our chosen parameterisation for an evolving IMF is derived from hydrodynamic star-forming cloud simulations.
Because of this self-regulation, the current observational constraints on the UV LF at $z\simeq12.5$ also remain slightly underpredicted for our evolving IMF model, even when assuming no attenuation of the UV light by dust. This implies that neither a top-heavier IMF nor the ejection of dust by radiative processes \citep{Ferrara2023} or a combination of the two can explain the early luminous galaxies alone, suggesting that these galaxies need to exhibit higher star formation efficiencies, which could be feedback-free starbursts \citet{Dekel2023}, or that our model underpredicts the stochasticity of star formation. Assuming the IMF becomes top-heavier towards higher redshifts, the higher abundance of massive stars could increase stochasticity in star formation due to their shorter lifetimes and stronger stellar feedback, affecting the surrounding ISM more dynamically. Hence, assuming that stars form constantly across a time step -- as we do in our model -- may not reflect this burstiness sufficiently.

Secondly, the lower average $f_\mathrm{esc}$ values in the evolving IMF model, paired with the increased gas-phase metallicity and possibly increased star formation efficiency, suggest that $z\gtrsim10$ galaxies may show strong Lyman-$\alpha$ and [O{\sc iii}] emissions. While these lower $f_\mathrm{esc}$ values may seem to be in contradiction with first expectations of a top heavier IMF leading to higher $f_\mathrm{esc}$ values due to the increased abundance of massive stars, they may reflect that the formation of fewer stars in the evolving IMF results in higher gas-to-stellar mass ratios in galaxies and fewer low-density tunnels through which ionising photons can escape. Potentially, the increased covering fraction of dense gas clouds and velocities of outflowing gas, resulting from a higher number of SNe exploding at a time, could cause a redshift in the Lyman-$\alpha$ line escaping from these early galaxies that is sufficient for a significant fraction of the Lyman-$\alpha$ radiation to be transmitted through the predominantly neutral IGM.
However, to fully understand how the formation of more massive stars at a slower rate changes $f_\mathrm{esc}$ and the Lyman-$\alpha$ line emerging from high-redshift galaxies, radiation hydrodynamic simulations and semi-analytic models are needed that follow how stellar feedback of stellar populations with different IMFs affect the structure and gas densities of the interstellar and circumgalactic media.

We note that, in theory, instead of reducing the star formation efficiency $f_\star$ in the evolving IMF model to fit the observed UV LFs, we could have assumed the same $f_\star$ value as in the Salpeter IMF model, essentially baselining our galaxy model parameters to the $z>10$ observed UV LFs, and adjust our dust model (i.e. assumptions on the dust radius and/or grain sizes) to reproduce the UV LFs at lower redshifts. However, with our current model, this option would result in a poorer fit at lower redshifts due to the $f_w$ and $f_\star$ values determining the position of the change in the UV LF slope that develops towards lower redshift (e.g. $M_\mathrm{UV}\simeq-18$ at $z=6$). The evolving IMF model's lower mass-to-light ratio would shift this change in slope to higher UV luminosities ($M_\mathrm{UV}\simeq-20$). Yet, galaxies fainter than the UV luminosity where the slope changes are predominantly SN feedback-limited exhibit minimal dust contents. Hence, aligning our model with observations in this scenario would require a dust model that suppresses the UV very efficiently in galaxies with a low dust content but only moderately in galaxies with higher dust contents, which would imply that the composition of dust and/or its concentration in the halo changes rapidly as galaxies grow in mass from $M_h\simeq10^{10.5}\msun$ to $10^{12}\msun$.

Furthermore, our upper mass limit of the IMF of $100\msun$ could be adjusted, given the observational evidence revealing the presence of very massive stars with masses exceeding $100\msun$ and sub-solar metallicities \citep[see e.g.,][]{Martins2023, Mestric2023, Wofford2023, Upadhyaya2024}. \citet{Martins2022} demonstrated that adding the spectra of very massive stars ranging from $100\msun$ to $300,\msun$ and with a metallicity of $Z=0.4Z_\odot$ enhances the ultraviolet continuum of a BPASS $0.1-100\msun$ spectra by approximately a factor two. 
Extending the mass range of our IMF models to include very massive stars would likely lead to a further decrease of the mass-to-light ratio, resulting in a lower star formation efficiency, $f_\star$, and lower ionising escape fraction normalisation, $f_\mathrm{esc}^0$. If we were to extend our current assumption -- where each star more massive than $8\msun$ explodes and injects an energy of $10^{51}$erg~s${-1}$ -- to stars with masses exceeding $100\msun$, an IMF including such very massive stars would yield a higher energy injected per stellar mass, compensating for at least some of the decrease in the mass-to-light ratio and driving the aforementioned self-regulation of galactic processes in the presence of an evolving IMF. 
However, this assumption might oversimplify the stellar evolution process. Stars with masses $\gtrsim 40\msun$ and lower metallicities $Z\lesssim10^{-2}-10^{-1}Z_{\odot}$ are expected to directly collapse into black holes \citep{Heger2003}. While most intermediate and massive galaxies in our simulation attain high enough metallicity values, this assumption may not hold for the lower-mass galaxies, where the energy injected by supernovae would be lower than currently assumed (allowing for a lower supernovae coupling efficiency, $f_w$). Yet, the evolution of metal-poor massive stars remains highly uncertain owing to uncertainties surrounding their mass loss, rotation, and binary interactions.

Observationally distinguishing between our IMF models poses challenges, given that common galaxy scaling relations often rely on stellar mass, whose inference depends on the assumed IMF. To address this, we need to explore IMF-independent observables or quantities derived from such observables. While the galaxy population's gas-phase metallicities show minor differences in their UV luminosity dependency between our IMF models, the UV luminosity dependencies of their dust masses differ increasingly towards higher redshifts. This suggests that combining dust mass estimates from ALMA observations with JWST UV measurements of the same galaxies could provide valuable constraints on the IMF during reionisation.

Our implementation of an evolving IMF into the {\sc astraeus} framework is a step towards a more-self-consistent model for galaxy evolution and reionisation and enables us now to further explore which observable galactic properties, such as various emission lines or UV slopes, and the relations among them, are best suited to constrain the nature of the IMF of the galaxies in the first billion years of evolution.


\begin{acknowledgements}
    We thank the anonymous referee for their comments which improved the quality of the paper.
    We also thank Koki Kakiichi and Viola Geli for useful discussions and comments.
    AH and CM acknowledge support by the VILLUM FONDEN under grant 37459. The Cosmic Dawn Center (DAWN) is funded by the Danish National Research Foundation under grant DNRF140. 
    PD and MT acknowledge support from the NWO grant 016.VIDI.189.162 (``ODIN"). PD warmly thanks the European Commission's and University of Groningen's CO-FUND Rosalind Franklin program. 
    CM also acknowledges support from the Carlsberg Foundation under grant CF22-1322.
    GY acknowledges  Ministerio de  Ciencia e Innovaci\'on (Spain) for partial financial support under research grant PID2021-122603NB-C21. 
    The {\sc vsmdpl} simulation has been performed at LRZ Munich within the project {\it pr87yi}. The authors gratefully acknowledge the Gauss Centre for Supercomputing e.V. (www.gauss-centre.eu) for funding this project by providing computing time on the GCS Supercomputer SUPERMUC-NG at Leibniz Supercomputing Centre (www.lrz.de). The CosmoSim data base (\url{www.cosmosim.org}) provides access to the simulation and the Rockstar data. The data base is a service by the Leibniz Institute for Astrophysics Potsdam (AIP). 
    This research made use of \texttt{matplotlib}, a Python library for publication quality graphics \citep{hunter2007}; and the Python library \texttt{numpy} \citep{numpy}.
\end{acknowledgements}

%
%

\bibliographystyle{aa} 
\bibliography{sources, hutter_papers, astraeus, observations, models, repositories} 

\begin{appendix} 

\section{Luminosities for continuous star formation}
\label{app_luminosities_contSF}

Equations.~\ref{eq:Lnu} and~\ref{eq:Q} describe the time evolution of the ionising emissivity and UV luminosity of a starburst for the evolving IMF. However, in {\sc astraeus} we assume that stars form continuously with a constant SFR over a time step. Here, we provide the expression we use in {\sc astraeus}. The ionising emissivity, $\dot{Q(t)}$, and UV luminosity, $L_\nu(t)$, of stars forming with a constant SFR $s_0$ from gas with metallicity $Z$ between times $t_i$ and $t_f$ at a given time $t\geq t_f > t_i$ and $z$ being the redshift at time $t_i$ are
{\small
\begin{eqnarray}
    \frac{\dot{Q}(t,t_i,t_f,s_0,z,Z}{\rm s^{-1}\msun^{-1}}&=& \frac{s_0}{\msun \rm yr^{-1}} 10^{e_0} \left( \left[ t_f - \mathrm{max}(t_i, \mathrm{min}[t-10^{6.35}\mathrm{yr},t_f]) \right]\right. \nonumber \\
    &-&\left.\frac{10^{-6.35 * e_1}}{e_1 + 1}\left[ (t-\mathrm{min}(t_f, \mathrm{max}[t-10^{6.35}\mathrm{yr},t_i]))^{e_1+1}\right.\right. \nonumber \\
    &-& \left.\left.(t-t_i)^{e_1+1} \right]  \right),
\end{eqnarray}
}%
with
{\small
\begin{eqnarray}
    e_0 &=& 45.93 - 1.17~(\lg Z + 2) ~ + ~ 0.064~z~(\lg Z + 3.46), \nonumber \\ 
    e_1 &=& -3.91 - z~\left[0.051~(\lg Z + 2.97) + 0.0239\right], \nonumber
\end{eqnarray}
}%
and
{\tiny
\begin{eqnarray}
    \frac{L_\nu(t, z, Z)}{\mathrm{erg}\mathrm{s}^{-1}\mathrm{Hz}^{-1}\msun^{-1}} &=& \frac{s_0}{\msun \rm yr^{-1}} * 10^{e_0} \left(  \left[ t_f - \mathrm{max}(t_i, \mathrm{min}[t-10^{6}\mathrm{yr},t_f]) \right]\right. \nonumber \\
    &-&\frac{10^{-6 * e_1}}{e_1 + 1}\left[ (t-\mathrm{min}(t_f, \mathrm{max}[t-10^{6}\mathrm{yr},t_i]))^{e_1+1}\right. \nonumber \\
    &-&\left.\left. (t-\mathrm{max}(t_i, \mathrm{min}[t-10^{6.45}\mathrm{yr},t_f]))^{e_1+1} \right]  \right) \nonumber \\
    &-&\frac{s_0}{\msun \rm yr^{-1}} * 10^{e_0+0.45e_1} \left(  \frac{10^{-6.45 * e_2}}{e_2 + 1}\right.  \\
    &\times& \left[ (t-\mathrm{min}(t_f, \mathrm{max}[t-10^{6.45}\mathrm{yr},t_i]))^{e_2+1}\right. \nonumber \\
    &-&\left.\left. (t-\mathrm{max}(t_i, \mathrm{min}[t-10^{8.5}\mathrm{yr},t_f]))^{e_2+1} \right]  \right) \nonumber \\
    &-&\frac{s_0}{\msun \rm yr^{-1}} * 10^{e_0+0.45e_1+2.05e_2} \left(  \frac{10^{-8.5 * e_3}}{e_3 + 1}\right.\nonumber \\
    &\times&\left.\left[ (t-\mathrm{min}(t_f, \mathrm{max}[t-10^{8.5}\mathrm{yr},t_i]))^{e_3+1}  - (t-t_i)^{e_3+1} \right]  \right), \nonumber
\end{eqnarray}
}%
with
{\small
\begin{eqnarray}
    e_0 &=& 31.947 - 0.921~(\lg Z + 1.676) ~ + ~ 0.0569~z~(\lg Z + 3.498), \nonumber \\
    e_1 &=& 0.6876, \nonumber \\
    e_2 &=& -1.56 - 0.0579~z, \nonumber \\
    e_3 &=& -2 + 0.03~z, \nonumber
\end{eqnarray}
}%
respectively.

\section{ SFR dependence of the stellar mass to metallicity relation}
\label{app_mass-metallicity-SFR}

\begin{figure*}
    \centering
    \includegraphics[width = \textwidth]{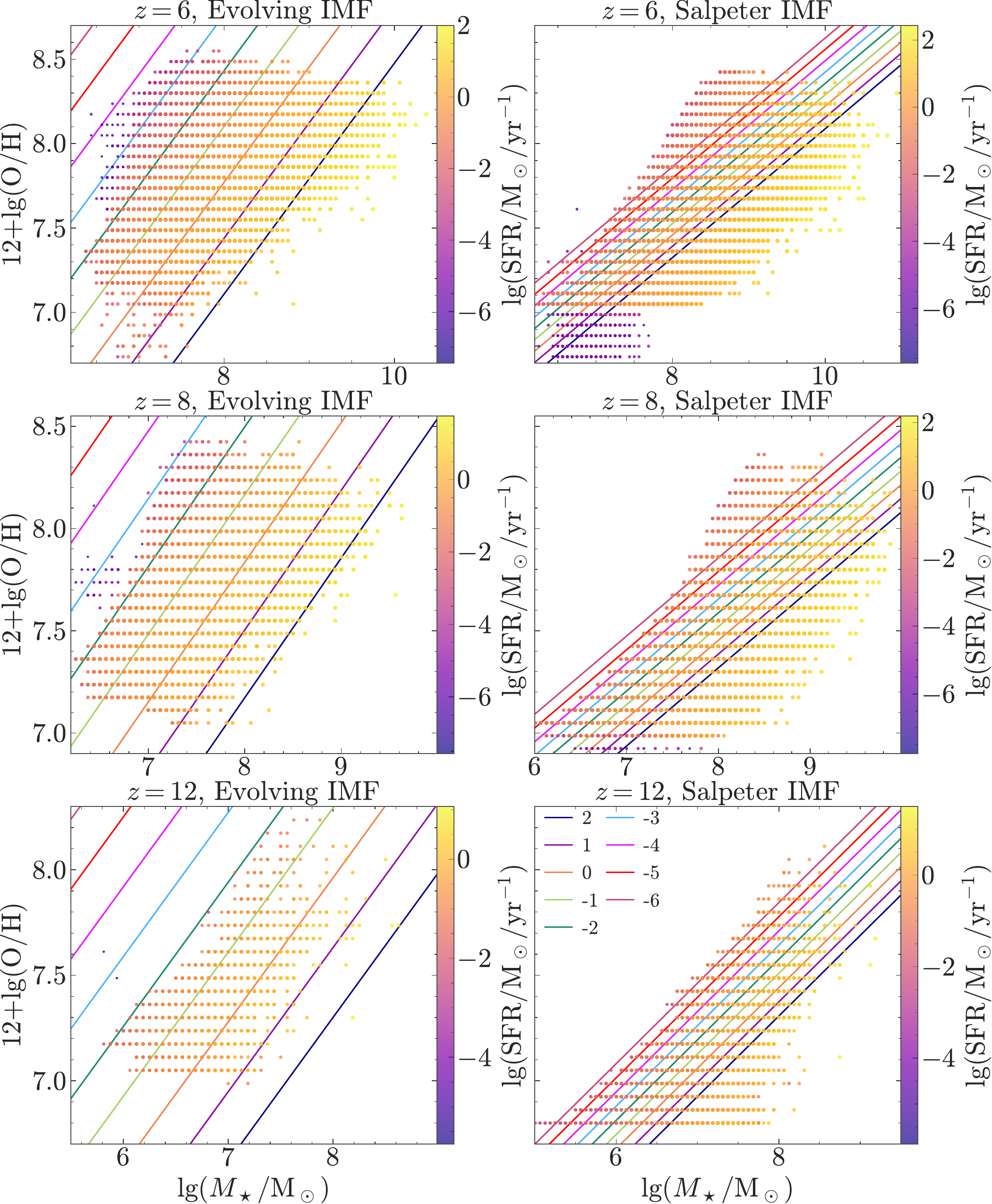}
    \caption{Best-fit high-redshift fundamental planes of metallicity (HFPZ) at $z=6$, $8$, $12$ for the evolving IMF (left) and Salpeter IMF (right) models. Coloured lines show the HFPZ at fixed SFR values, while the tiles depict the median SFR values in the respective stellar mass -- metallicity bin according with their colours encoding the SFR value according to the colour bar on the right side of each panel.}
    \label{fig:HFPZ}
\end{figure*}

In Fig.\ref{fig:HFPZ}, we present the stellar mass-metallicity relations at $z=6$, $8$, and $12$ for both the evolving IMF (left) and the Salpeter IMF (right) models. In contrast to Fig.\ref{fig:Z_Mstar}, which illustrates the number distribution of galaxies in the stellar–metallicity plane along with the resulting median relations, this figure showcases (i) the median SFR value in each stellar mass–metallicity bin as coloured tiles and (ii) the high-redshift fundamental plane of metallicity (HFPZ) at different SFR values ranging from $\lg(\mathrm{SFR})=-6$ to $2$.
We note two differences in the HFPZs of the evolving IMF and Salpeter IMF models: Firstly, the evolving IMF exhibits less scatter in SFR values across the MZR than the Salpeter IMF at all redshifts. This reduced scatter can be attributed to (i) the $2.5\times$ lower star formation efficiency for more massive galaxies and (ii) a higher fraction of the stars forming exploding as SNe in the same time step (and thus reducing the SFR in that time step), as the IMF is more top-heavy, leading to massive stars exploding earlier (see Sect. \ref{subsec_starformation_main_sequence} for the detailed explanation).
Secondly, the SFR for a fixed metallicity and stellar mass is lower in the evolving IMF than the Salpeter IMF model due to its lower mass-to-light ratio. This lower mass-to-light ratio also results in a weaker correlation between metallicity and SFR.

\section{ Intrinsic and escaping ionising emissivities}
\label{app_ionising_emissivities}

\begin{figure*}
    \begin{subfigure}{0.5\textwidth}
        \centering
        \includegraphics[width = \textwidth]{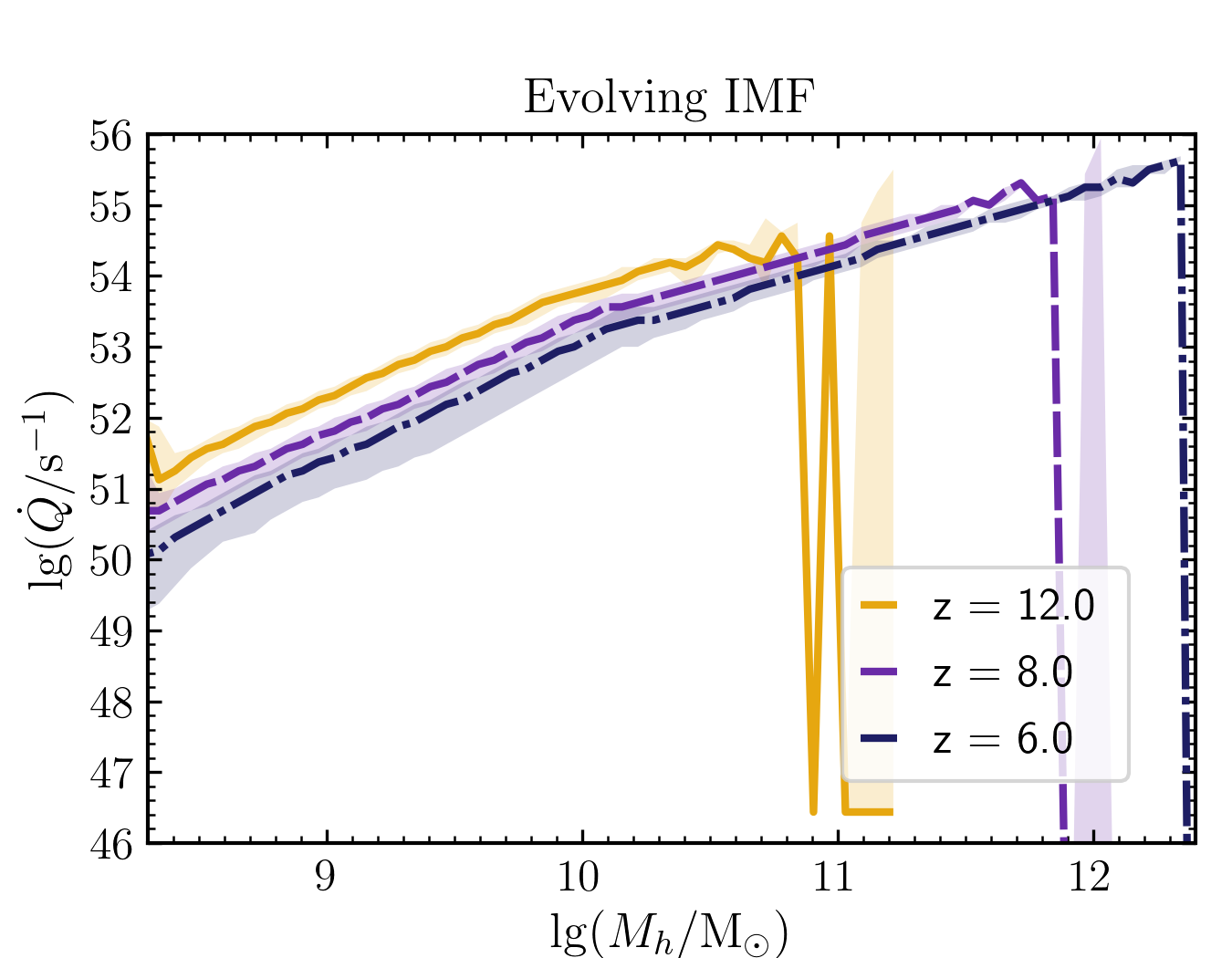}
        \caption{Intrinsic ionising emissivity for the evolving IMF model}
        \label{fig:Q_evolvingIMF}
    \end{subfigure}
    \begin{subfigure}{0.5\textwidth}
        \centering
        \includegraphics[width = \textwidth]{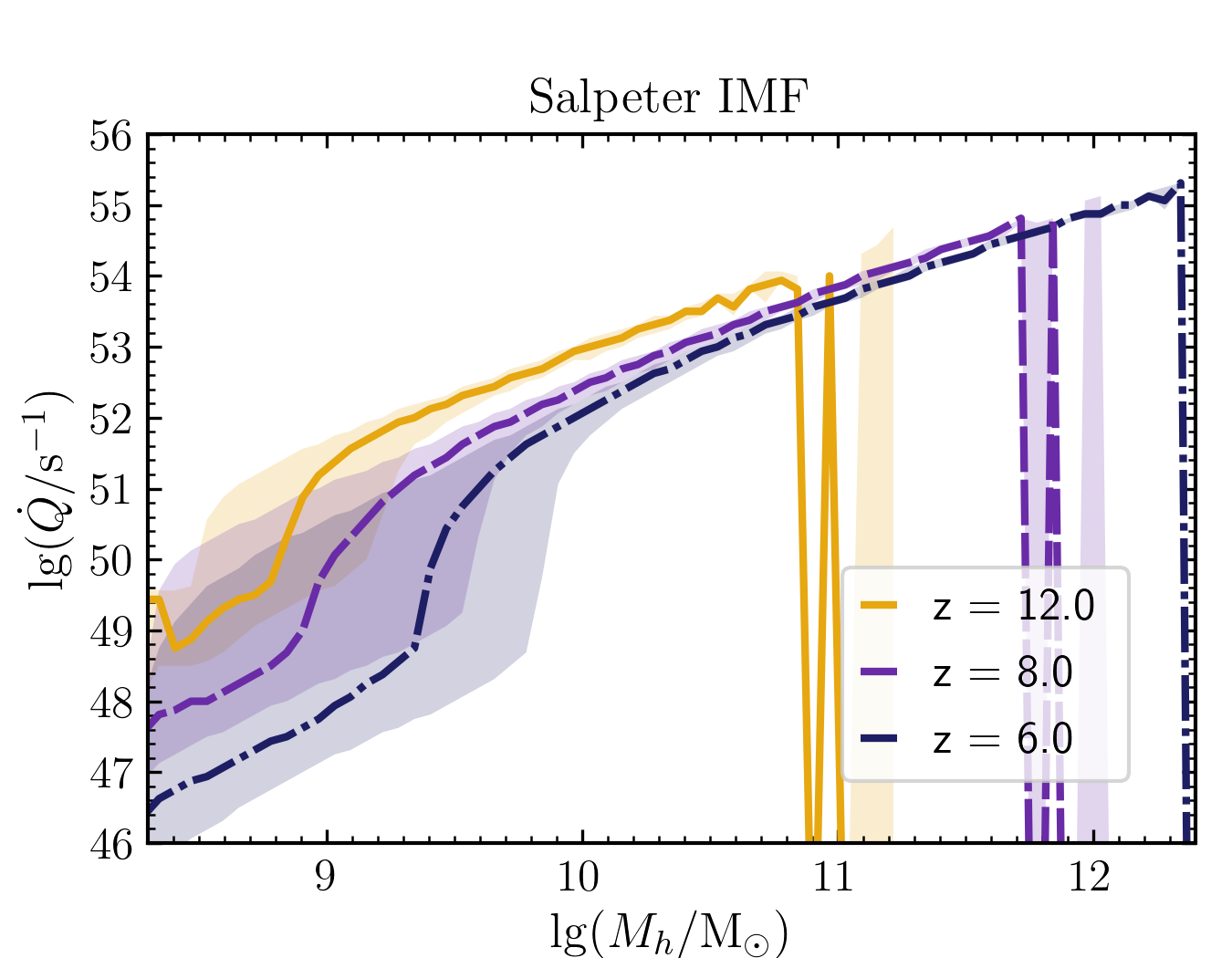}
        \caption{Intrinsic ionising emissivity for the Salpeter IMF model}
        \label{fig:Q_SalpeterIMF}
    \end{subfigure}
    \begin{subfigure}{0.5\textwidth}
        \centering
        \includegraphics[width = \textwidth]{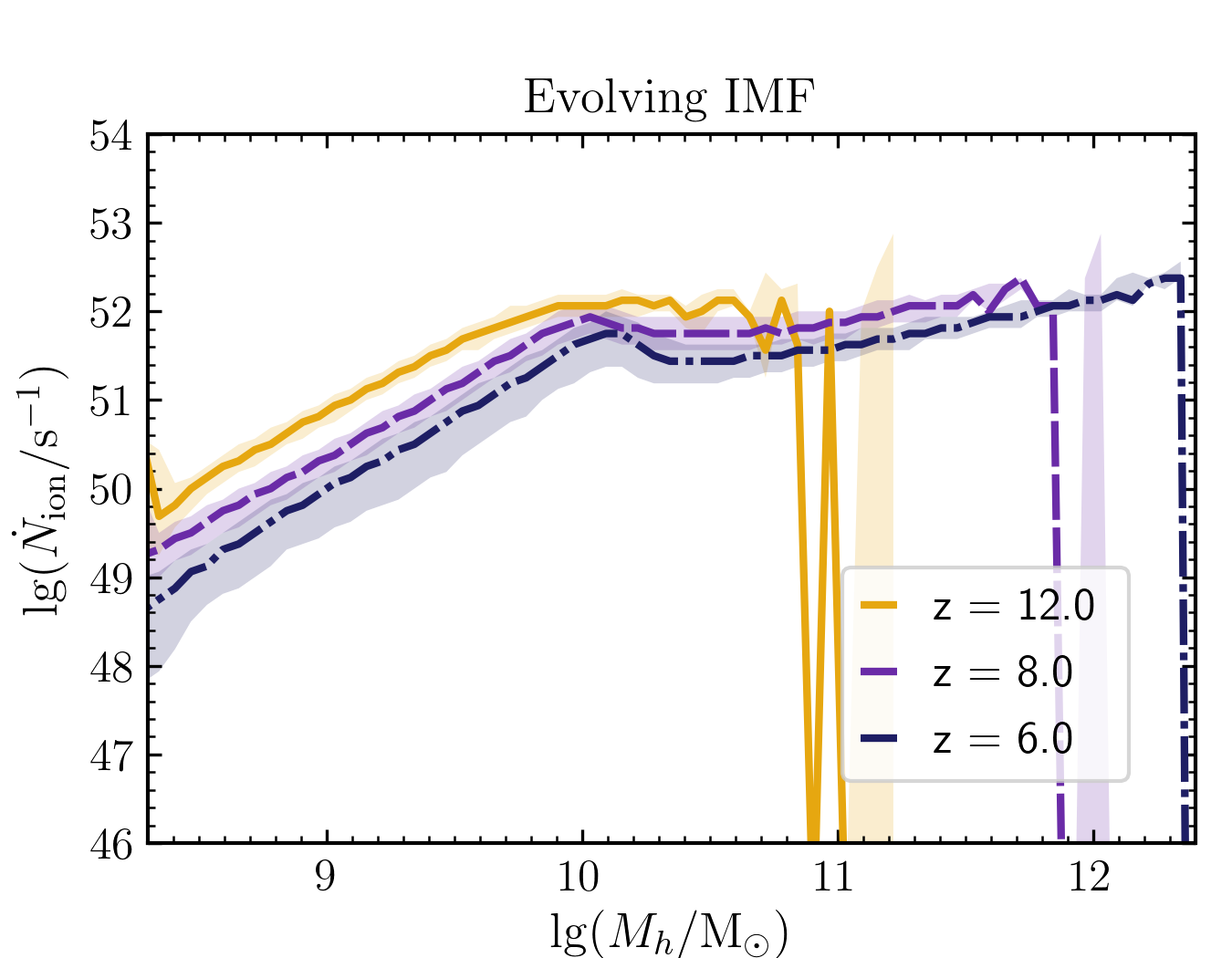}
        \caption{Escaping ionising emissivity for the evolving IMF model}
        \label{fig:Nion_evolvingIMF}
    \end{subfigure}
    \begin{subfigure}{0.5\textwidth}
        \centering
        \includegraphics[width = \textwidth]{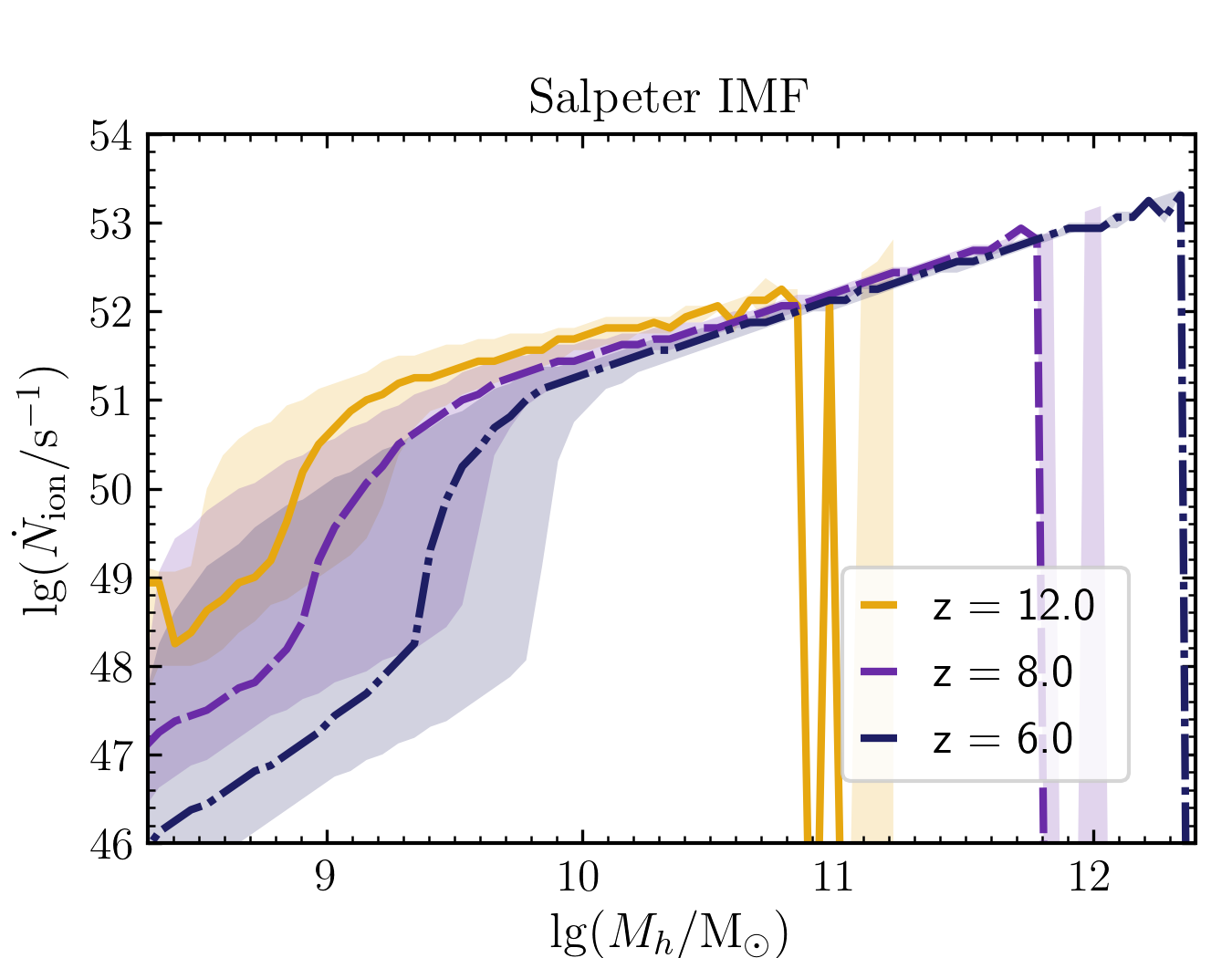}
        \caption{Escaping ionising emissivity for the Salpeter IMF model}
        \label{fig:Nion_SalpeterIMF}
    \end{subfigure}
    \caption{Intrinsic (top) and escaping (bottom) ionising emissivities as a function of halo mass $M_h$ for the evolving IMF (left) and Salpeter IMF (right) models at $z=6$ (dark blue), $8$ (violet), and $12$ (orange). Solid lines show the median ionising emissivity and transparent shaded coloured regions the $1-\sigma$ limits of the ionising emissivity values at a given halo mass.}
    \label{fig:ionising_emissivities}
\end{figure*}

In Fig.~\ref{fig:ionising_emissivities}, we show the distribution of intrinsic ($\mathrm{\dot{Q}}$, top panels) and escaping ionising emissivities ($\dot{N}_\mathrm{ion}=f_\mathrm{esc}\dot{Q}$, bottom panels) as a function of halo mass for the evolving IMF (left) and Salpeter IMF (right) models. We note the following differences between the two IMF models:
Firstly, due to its lower mass-to-light ratio, the evolving IMF model's intrinsic ionising emissivity is always higher than the one in the Salpeter IMF model (see Sect. \ref{subsubsec_light_to_mass_ratio}).
Secondly, the drop of the ionising escape fraction, $f_\mathrm{esc}$, towards lower stellar masses occurs at higher stellar masses in the evolving IMF than in the Salpeter IMF model. This results from the evolving IMF model's stronger SN feedback, expelling a larger fraction of gas in more massive galaxies.
Thirdly, the evolving IMF model's relation between the median escaping ionising emissivity and halo mass is flatter (i.e. low-mass galaxies ($M_{h}\lesssim10^{10}\msun$) have larger escaping ionising emissivity values, while more massive galaxies ($M_{h}\gtrsim10^{10}\msun$) exhibit lower ones) and has less scatter for low-mass galaxies ($M_{h}\lesssim10^{10}\msun$) than in the Salpeter IMF model. These differences arise from the more immediate nature of SN feedback in the evolving IMF model. The more immediate feedback is attributed to the fact that more stars that explode as SNe are more massive and have shorter lifetimes. Consequently, SN feedback in the evolving IMF regulates star formation efficiency within the same time step. In contrast, the Salpeter IMF model experiences a broader range of star formation efficiencies because more of the energy released by SN explosions originates from stars in previous time steps.

\end{appendix}

\end{document}